\documentclass[11pt,a4paper,fleqn]{article}
\usepackage[utf8]{inputenc}
\usepackage{authblk}
\usepackage{geometry}
\usepackage{xcolor}
\usepackage[longnamesfirst]{natbib}
\usepackage{graphicx}
\usepackage{pdflscape}  
\usepackage{appendix}
\usepackage{threeparttable}
\usepackage{booktabs}

\usepackage{amssymb}
\usepackage{amsbsy}
\usepackage{bm}
\usepackage{amsmath}
\usepackage{amsthm}
\usepackage{graphicx}
\usepackage{mathtools}
\usepackage{rotating}
\usepackage{setspace}
\usepackage{color, colortbl}
\definecolor{ashgrey}{rgb}{0.7, 0.75, 0.71}
\definecolor{columbiablue}{rgb}{0.61, 0.87, 1.0}
\definecolor{coral}{rgb}{1.0, 0.5, 0.31}

\definecolor{colBVAR}{HTML}{bababa}
\definecolor{colBART}{HTML}{d7191c}
\definecolor{colmixBART}{HTML}{fdae61}
\definecolor{colerrorBART}{HTML}{abd9e9}
\definecolor{colfullBART}{HTML}{2c7bb6}

\definecolor{colcons}{HTML}{e31a1c}
\definecolor{colSV}{HTML}{a6cee3}
\definecolor{colhBART}{HTML}{1f78b4}

\usepackage{enumitem}
\newlist{steps}{enumerate}{1}
\setlist[steps,1]{label = Step \arabic*:}

\usepackage{adjustbox}
\usepackage{dcolumn}
\newcolumntype{d}[1]{D..{#1}} % for alignment of numbers on decimal marker

\usepackage{caption}
\usepackage{subcaption}
\definecolor{nblue}{HTML}{000660}
\usepackage[colorlinks=true,urlcolor=nblue,linkcolor=nblue,citecolor=nblue]{hyperref}

\newcommand*{\myeqref}[2][Eq.~]{%
  \hyperref[{#2}]{#1(\ref*{#2})}%
}
\def\equationautorefname#1#2\null{%
  Eq.#1(#2\null)%
}

\begin{document}
\title{\textbf{Measuring Shocks to Central Bank Independence using Legal Rulings}\thanks{
%\textit{Corresponding author}: Name. Affiliation. \textit{Address}: Address. \textit{Email}: \href{mailto:name@something.com}{name@something.com}. 
The authors are grateful to Marek Jaroci\'nski for providing many helpful suggestions and parts of the dataset. Huber and Pfarrhofer gratefully acknowledge financial support from the Austrian Science Fund (FWF, grant no. ZK 35) and from the Oesterreichische Nationalbank (OeNB, grant no. 18304).}}

\author[a]{Stefan \textsc{Griller}}
\author[b]{Florian \textsc{Huber}}
\author[b]{Michael \textsc{Pfarrhofer}}
\affil[a]{\textit{Vienna University of Economics and Business}}
\affil[b]{\textit{University of Salzburg}}

\date{\today}

\maketitle\thispagestyle{empty}\normalsize\vspace*{-2em}\small\linespread{1.5}
\begin{center}
\begin{minipage}{0.8\textwidth}
%\noindent\small In this paper we investigate the empirical consequences of legal rulings on the conduct of monetary policy. The unconventional monetary policies of the European Central Bank have come under legal scrutiny in national courts and the European Court of Justice. These lawsuits (and their subsequent outcomes) have the potential to severely impact the scope and flexibility of central bank policies and central bank independence in a wide sense, with important consequences for the European economy and financial markets. Since the number of events is relatively small, we develop an econometric approach that searches for minimum variance observations which we use to isolate the effects of ruling announcements. Our results suggest that legal rulings have a powerful effect on financial markets. Expansionary ruling shocks increase financial market stability and stock prices, decrease government bond yields relative to German bonds as well as decrease corporate bond spreads. Inflation swap reactions suggest higher levels of inflation, with stronger effects on the short-end of the inflation swap curve. % 164 words
\noindent\small We investigate the consequences of legal rulings on the conduct of monetary policy. Several unconventional monetary policy measures of the European Central Bank have come under scrutiny before national courts and the European Court of Justice. These lawsuits have the potential to severely impact the scope and flexibility of central bank policies, and central bank independence in a wide sense, with important consequences for the real and financial economy. Since the number of relevant legal challenges is small, we develop an econometric approach that searches for minimum variance regimes which we use to isolate and measure the effects of these events. Our results suggest that legal rulings addressing central bank policies have a powerful effect on financial markets. Expansionary shocks ease financial conditions along various dimensions, and inflation swap reactions suggest inflationary pressures with stronger effects in the short term.

factor model, identification through heteroskedasticity
\end{minipage}
\end{center}\vspace*{3em}

\normalsize\clearpage
\section{Introduction}
The advantages  of and challenges to an independent central bank have been emphasized in a broad body of academic papers  over the last 30 years \citep[see, e.g.,][]{alesina1988macroeconomics, cukierman1992measuring,cukierman1994central, alesina1993central,fischer1995central,crowe2007evolution,walsh2011central, binder2021political}. {In many economies, legislators have been improving the respective legal architectures and procedures to make central banks less reliant on day-to-day policy making and to free them from undue political pressures and influence. While measuring the precise economic consequences of such changes is difficult, there exists some evidence that related shifts in central bank policies in the United States (US) and the United Kingdom (UK) have led to declining levels of inflation, decreasing inflation volatility and the stabilization of inflation expectations \citep{cukierman1986theory,blinder2008central}.}

During the global financial crisis and the Covid-19 pandemic, central banks such as the US Federal Reserve and the Bank of England, however, actively cooperated with governments to fight these economic downturns. Such coordinated action did not harm central bank independence due to the legal structure of the Bank of England and the Federal Reserve System. In the euro area, coordinated actions also took place, but with member states exhibiting heterogeneous macroeconomic and, most importantly, debt dynamics. The question whether the European Central Bank (ECB) is legally allowed to intervene frequently in this regard arose in the media and policy discussions.

The independence of the ECB constitutes a main pillar of the design of the European Monetary Union (EMU). In the last decade, ECB expertise and capacity have been challenged by several severe economic crises which called for unconventional monetary policy actions \citep[see][for a review]{hartmann2018european}. These actions involved buying government debt securities in unprecedented amounts and this raised the question whether the ECB is exceeding its mandate. As a result, selected ECB programs have come under legal scrutiny. These lawsuits on the (il)legality of the ECBs unconventional monetary policy measures could, in the worst case, severely limit the scope of the ECB's options during crises and thus also its independence and credibility in the context of timely and effective policy making. This argument gives rise to our interpretation of the outcome of such a ruling as a central bank independence shock.\footnote{For a recent study that uses Twitter data to identify central bank independence shocks, see \cite{bianchi2019threats}.}

We take a closer look at three preliminary rulings handed down by the European Court of Justice (CJEU) at the request of highest national courts, and at the related decisions of these courts. We scrutinise the Pringle Case which concerned the legality of establishing the European Stability Mechanism (ESM) which is a member states’ measure, and was brought up by the Irish Supreme Court; further, we assess the Gauweiler and the Weiss Case, both raised by the German Federal Constitutional Court (GFCC), and both concerning non-traditional ECB measures which allegedly were outside the ECB’s mandate.

In this paper, we aim to exploit variation surrounding the announcements of these rulings to identify a central bank independence shock and evaluate its effect on a broad panel of crucial financial indicators. We achieve identification by assuming that the variance of the endogenous variables is higher on ruling days than on so-called non-event days.\footnote{This feature of exploiting heteroskedasticity for shock identification relates our paper to a broad literature on using time-varying volatilities in macroeconomic and financial data to pin down structural shocks statistically for enabling economically meaningful interpretation \citep[see, e.g.,][]{sentana2001identification,rigobon2003identification,lewis2017robust,lewis2021identifying,bertsche2020identification,brunnermeier2021feedbacks}.} Our approach is closely related to the factor model proposed in \cite{gurkaynak2020missing}.  Since the number of events (i.e., days with news about rulings) relative to the non-event days is very small, we propose a mixture model that endogenously selects the non-event days. This mixture model assumes that the panel of time series features a common volatility process moving through a set of regimes \citep[see also][]{carriero2016common}. The non-event set is then defined to comprise of observations with minimum variance, and thus captures the notion that if the common variance is minimal, we can safely regard this regime as a non-event regime (i.e., a regime without meaningful events that could jeopardize our identification strategy).

{We consider a panel of time series including euro area-wide, country-level and international information. Among several key variables of interest, we control for cross-country movements in 10-year yields by including government bond yields relative to German Bunds for several euro area member states. These series are complemented by financial indicators such as measures of systemic stress, short-term interest rates, corporate bond spreads, as well as stock market and exchange rate information. Inflation expectations are captured by including inflation swaps at different maturities, and oil and gold prices are included as important non-domestic variables. To estimate the effects of independence shocks on this panel of series we compute local projections within a unified econometric framework and thus obtain a dynamic picture on the effects of these rulings.}

Our results can be summarized as follows. A positive independence shock (i.e., no impairment in the variety of policy actions available to the central bank) leads to lower spreads between German 10-year Bunds and 10-year government bond yields across selected euro area member states. Considering the cross-sectional responses reveals appreciable heterogeneity with respect to the dynamic responses. Countries such as Greece, Spain and Italy display more pronounced reactions whereas countries located to the euro area core seem to be somewhat less affected. {A similar pattern occurs for measures of financial stress across euro area member states. On aggregate, expansionary shocks ease financial conditions, yield gains on stock markets and result in modest increases in short-term inflation expectations.}

The paper proceeds as follows. Section \ref{sec: rulings} provides the legal background and a brief summary of the three major legal cases which we use to form our set of events. Section \ref{sec: econometrics} introduces the empirical framework used to estimate the effects of the rulings in our panel of time series. Section \ref{sec: emp_results} presents the main empirical results of the paper. The final section summarizes and concludes the paper.

\section{Controversies on delimiting monetary and economic policy, and resulting litigation} \label{sec: rulings}
\subsection{Exposition of the fundamental legal dimensions}
ECB independence is a cornerstone of the EMU design -- it is triple guaranteed by primary law: by Articles 130 and 282(3) Treaty on the Functioning of the European Union (TFEU), and Article 7 Protocol Number 4 on the Statute of the European System of Central Banks and of the European Central Bank. Moreover, ECB independence is bolstered by flanking provisions, among those the prohibition on monetary financing (Article 123 TFEU), and the appointment mechanism as well as the not renewable eight-years term of office for members of the Executive Board. The underlying rationale is that delivering on the ECB's primary objective, price stability (Article 127 TFEU), would be made more difficult if not endangered by external, not the least political influence of any kind.

ECB independence is guaranteed by the Treaties and thus strongly protected under European Union (EU) law not only against undue influence on ECB organs but also against encroachments by member states' and other EU institutions' activities. Any violation can be brought before the CJEU (Articles 260 and 263 TFEU). For ``monetary policy for the Member States whose currency is the euro'' is an exclusive competence of the EU, with the ECB as the EU institution entrusted to take care of; ``only the Union may legislate and adopt legally binding acts, the Member States being able to do so themselves only if so empowered by the Union or for the implementation of Union acts'' (Articles 2(1) and 3(1)(c) TFEU).

Independence, however, does not exempt the ECB from the EU legal order. The ECB is subject to legal scrutiny by the CJEU. The latter emphasised early on that independence ``does not have the consequence of separating'' the ECB from every rule of EU law (CJEU, Case C-11/00, \S135). First and before anything else, the ECB must act within the limits of the powers conferred on it (Article 13(2) Treaty on European Union, TEU), which is to ``conduct the monetary policy of the Union'' (Article 282(1) TFEU). Acting outside that mandate, or more generally, violating EU law, would render ECB measures voidable by the CJEU.

Both aspects -- the member states' respect for the ECB’s powers and, inversely, the ECB's respect for its own limits which protect the member states' economic (including fiscal) policy powers which they did not transfer to but merely ``coordinate'' within the Union (Article 5(1) TFEU) -- became controversial during and in the aftermath of the economic and financial crisis starting in $2008$. The controversy led to several high-profile judgments clarifying the limits between European \textit{economic} and \textit{monetary} policy. The cases will be briefly sketched out below \citep[for a more detailed analysis, see][]{griller2021analytical}. All three of them had been filed with the CJEU by highest national courts as preliminary ruling procedures (Article 267 TFEU), all of them putting critical questions concerning the legality of measures taken. Only the second and third judgment (Gauweiler and Weiss) directly concerned (unconventional) ECB measures, the first (Pringle) was on a measure taken by the member states. However, all three of them dealt with the scope of monetary policy and thus ECB independence, and can therefore, if only in a wide sense regarding Pringle, be addressed as possible independence shocks.\footnote{The ECB's Outright Monetary Transactions (OMT) program is conditional on a memorandum of understanding between a member state and the EMS, so the Pringle case had the potential to nullify the OMT, thus directly impacting the ECB's toolkit.} Table \ref{tab:rulingdates} provides a list and further details.\footnote{We have verified that the dates of the rulings do not systematically coincide with other important news. In particular, we have cross-checked with the dates of the following press releases: Eurostat releases of euro area flash CPI, core CPI, produces prices, M3 money growth, trade balance, unemployment rate, business and consumer surveys, national accounts, industrial production, retail sales; Markit euro area PMI indices; ECB monetary policy announcements, speeches by ECB Presidents; Federal Reserve Bank monetary policy announcements (scheduled and unscheduled FOMC press releases); US initial jobless claims and non-farm payrolls. For most releases the overlap is on less than three dates, and often on zero dates.}

The outcomes had been far from obvious. The CJEU's judgments included a certain margin of discretion. Indeed, all these judgments have been controversial before, during and in the aftermath of litigation, both at CJEU and national level. To put it differently, \textit{ex ante}, these lawsuits had the potential of ousting crisis intervention measures as illegal. As a result, these measures would have had to be terminated. That would have caused a serious challenge to both the member states' and the ECB's efforts to contain the respective crises and to counteract their adverse effects. Thus, stock market and interest rate reactions to these decisions are specifically pertinent when it comes to measuring the economic relevance of court decisions. 

The reactions reflect the mentioned uncertainty and possible expectations or suspicions of a different outcome. If the court decision as such comes as a surprise to the markets, which in our cases is conceivable mainly for the national court decision to request the preliminary ruling by the CJEU (see below), it is this unexpected event which may be considered as a shock that is provoking market reactions. Identifying the details of these deviating expectations and their influence on the resulting dimension of market reactions -- including the important differences between different litigation stages -- goes beyond the scope of our study and might be worthwhile future research. However, this would necessitate digging both into the economic weight of the matter and the quality of \textit{ex ante} market expectations \textit{vis-\`a-vis} the respective judgments, posing a challenge to any research design.

Preliminary ruling procedures under Article 267 TFEU normally provide at least four, and sometimes even more, prominent stages which are worthwhile examining. First, the decision of the national court to interrupt its proceedings and ask the CJEU for an answer to a difficult question of interpretation or on the validity of secondary EU law. Second, the (non-binding) Opinion of the CJEU's Advocate General (AG) whereby the latter proposes how the CJEU should decide. Third, the requested answer of the CJEU given in its judgment which is binding under EU law. Fourth, the judgment or decision of the national court based on the answer given by the CJEU.

\begin{table}[t]
\caption{Ruling dates and description}\label{tab:rulingdates}\vspace*{-0.5em}
\centering\small
\begin{threeparttable}
\begin{tabular}{p{0.3cm}p{1.8cm}p{2.2cm}p{2.5cm}p{2cm}p{4.75cm}}
\toprule
& \textbf{Decision} & \textbf{Publication} & \textbf{Number} & \textbf{Court} & \textbf{Description} \\ 
  \midrule
  \multicolumn{6}{l}{\textit{Case C-370 (Pringle), ECLI:EU:C:2012:756}} \\
  \cmidrule(lr){1-6}
 & 2012-07-31 & 2012-07-31 &  [2012] IESC 47 & SC, Ireland & Request for preliminary ruling\\ % $-$
 % & 2012-10-04 &  & C-317/12 & CJEU, president & Order accelerated procedure \\
 & 2012-10-26 & 2012-10-26 & C-317/12 & CJEU, AG & Opinion of the AG \\ % $+$
 & 2012-11-27 & 2012-11-27 & C-317/12 & CJEU & Judgement \\ % $+$
 % & 2014-04-04 & & [2014] IEHC 174 & HC & Decision on costs; information that plaintiff had repealed his appeal \\
  \midrule
  \multicolumn{6}{l}{\textit{Case C-62/14 (Gauweiler), ECLI:EU:C:2015:400}} \\
  \cmidrule(lr){1-6}
 & 2014-01-14 & 2014-02-07 & 2 BvR 2728/13 & GFCC & Request for preliminary ruling \\ % $-$
 & 2015-01-14 & 2015-01-14 & C62/14 & CJEU, AG & Opinion of the AG \\ % $+$
 & 2015-06-16 & 2015-06-16 & C62/14 & CJEU & Judgement \\ % $+$
 & 2016-06-21 & 2016-06-21 & 2 BvR 2728/13 & GFCC & Judgement \\ % $+$
 & & & 2 BvR 2729/13 & & \\
 & & & 2 BvR 2730/13 & & \\
 & & & 2 BvR 2731/13 & & \\
 & & & 2 BvE 13/13 & & \\
  \midrule
  \multicolumn{6}{l}{\textit{Case C-493/17 (Weiss), ECLI:EU:C:2018:1000}} \\
  \cmidrule(lr){1-6}
 & 2017-07-18 & 2017-08-15 & 2 BvR 859/15& GFCC & Request for preliminary ruling \\ % $-$
 & & & 2 BvR 1651/15 & & \\
 & & & 2 BvR 2006/15 & & \\
 & & & 2 BvR 980/16 & & \\
 & 2018-10-04 & 2018-10-04 & C-493/17 & CJEU, AG & Opinion of the AG \\ % $+$
 & 2018-12-11 & 2018-12-11 & C-493/17 & CJEU & Judgement \\ % $+$
 & 2020-05-05 & 2020-05-05 & 2 BvR 859/15 & GFCC & Judgement \\ % $-$
 & & & 2 BvR 1651/15 & & \\
 & & & 2 BvR 2006/15 & & \\
 & & & 2 BvR 980/16 & & \\
%$+$ & 2021-04-29 & 2021-05-18 & 2 BvR 161/15 & GFCC & Order of execution, dismissal \\ % not in sample
% & & & 2 BvR 2006/15 & & \\
%$-$ & 2021-06-08 & 2021-06-08 & Letter & EC & Letter of formal notice under Article $258$ TFEU \\ % not in sample
  \bottomrule
\end{tabular}
\begin{tablenotes}[para,flushleft]
\scriptsize{\textit{Notes}: Abbreviations are Advocate General (AG), Court of Justice of the European Union (CJEU), European Commission (EC), German Federal Constitutional Court (GFCC), Supreme Court (SC). {In our empirical work, we consider publication dates in case they differ from the decision dates for the timing of the shocks. Publication dates are those of relevance, since this is when the public is informed about the respective decision and judges are prohibited by law from disclosing their rulings prematurely.}}
\end{tablenotes}
\end{threeparttable}
\end{table}

\subsection{The Pringle case: Challenging the European Stability Mechanism}
In this case, the legality of the Treaty Establishing the European Stability Mechanism (ESM Treaty) which is a treaty concluded under international law by all EMU member states was under scrutiny. Operational from 2012, the ESM is (primarily) entitled to grant loans to euro area member states in economic difficulties, conditional on performing a macroeconomic adjustment program. The overall condition is that such support is necessary to safeguard the financial stability of the euro area as a whole and of ESM members. The ESM's maximum lending capacity is EUR $500$ billion, based on EUR $80.5$ billion paid-in capital by its member states, and EUR $724.3$ billion of committed callable capital.

The reason for the Irish challenge was, not the least, an effort to avoid the financial burden connected with ESM membership. It was claimed that Ireland (\textit{pars pro toto} of all member states), by ratifying the Treaty, would violate several provisions of the TEU and the TFEU on EMU and would directly affect the respective exclusive EU competence. Most and before all, there was the alleged violation of the so-called no bailout clause (Article 125 TFEU).

Both arguments were rejected by the CJEU (Case C-370/12, Pringle). The first was done based on the Court's definition, for the first time in such clarity, of monetary policy by its objectives -- mainly price stability -- rather than its instruments. The purpose of the ESM, as the Court stressed, is not to maintain price stability, but rather meeting the financial requirements of members facing severe financing problems: ``Even if the activities of the ESM might influence the rate of inflation, such an influence would constitute only the indirect consequence of the economic policy measures adopted'' (\S97). Consequently, the CJEU found EU competences to be respected.

The Court saw Article 125 TFEU respected based on his interpretation that this provision ``is not intended to prohibit either the Union or the Member States from granting any form of financial assistance whatever to another Member State'' (\S130). Article 125 TFEU requires that the member states remain subject to the logic of the market when entering into debts (\S135); ``[...] it must be held that that provision prohibits the Union and the Member States from granting financial assistance as a result of which the incentive of the recipient member state to conduct a sound budgetary policy is diminished'' (\S136). Financial assistance remains possible as long as the recipient member state remains responsible for its commitments, and the conditionality prompts a sound budgetary policy. To some surprise, ESM assistance was found compatible with these requirements -- given that the ESM does not act as a guarantor of the debts and does not assume them: ``On the contrary, such assistance amounts to the creation of a new debt, owed to the ESM by that recipient member state, which remains responsible for its commitments to its creditors in respect of its existing debts'' (\S139).

Accepting the ESM as a measure not encroaching on the EU's exclusive monetary union competence and thus respecting the ECB's independence was less spectacular than denying any violation of Article 125 TFEU, the so-called no bailout clause \citep[see, e.g.,][]{craig2013pringle,ohler2013kommentar}. It was this aspect which made the case so controversial which could not escape the attention of financial markets. The danger that a CJEU verdict to the contrary would mean the end of the ESM had been imminent.

\subsection{The Gauweiler case: The first challenge to the ECB's non-traditional measures before the CJEU}
In 2012 the ECB announced its readiness to perform Outright Monetary Transactions (OMT) -- in essence, purchases in secondary sovereign bond markets aiming at safeguarding an appropriate monetary policy transmission and the singleness of the monetary policy. The program was considered for euro area member states under an ESM adjustment program, and subject to conditionality. The announcement was accompanied by a statement from ECB president Mario Draghi: `` We will do whatever it takes within our mandate [...] to have a single monetary policy [...], to maintain price stability [...] and to preserve the euro.'' The program itself never became operative.

Nevertheless, it was challenged before the GFCC. The latter in turn introduced its first ever preliminary ruling procedure with the CJEU, clearly insinuating that the ECB's announcement might be beyond its mandate (\textit{ultra vires}), for resulting in an economic policy measure instead of respecting the limits of monetary policy as required under the treaties.

The CJEU dismissed the argument and upheld the ECB's announcement as legal. It confirmed its findings in Pringle that determining monetary policy must be done by referring ``principally to the objectives of that measure.'' Instruments are ``also relevant,'' only (Case C-62/14, Gauweiler, \S46). Safeguarding the singleness and an appropriate transmission of monetary policy would likely contribute to the primary objective of price stability (\S\S47–50). ``Indeed, a monetary policy measure cannot be treated as equivalent to an economic policy measure merely because it may have indirect effects on the stability of the euro area [...]'' (\S52). As a result, purchasing government bonds on the secondary market might be both: economic policy if done by the ESM, but monetary policy if done by the ECB (\S64). 

A different outcome would have invalidated the OMT program, resulting in a prohibition to uphold or renew it. The GFCC accepted, if somehow reluctantly, the verdict of the CJEU, and consequently dismissed and rejected the constitutional complaints before it (Case 2 BvR 2728/13 \textit{et al.}, June 21, 2016).

\subsection{The Weiss case: A legal worst-case scenario turning reality}
The second ever preliminary ruling procedure introduced by the GFCC also concerned an ECB program, equally insinuating that it might be \textit{ultra vires}. At stake was the ``Public Sector Purchase Programme'' (PSPP), the most important part of the ECB's non-traditional measures taken to fight the economic crisis, also labeled quantitative easing (QE). Starting in 2015, the ECB purchased sovereign bonds in secondary markets up to EUR $80$ billion per month (2016 and 2017) in order to support economic stability and avert deflationary pressures.

Again, however, the CJEU denied that the ECB had overstepped its mandate and upheld the program. It did so by reconfirming that determining whether a measure falls within the area of monetary policy is principally depending upon the objectives pursued, while the instruments employed are only of secondary relevance (Case C 493/17, Weiss, \S\S53ff). Equally it reconfirmed ``that a monetary policy measure cannot be treated as equivalent to an economic policy measure for the sole reason that it may have indirect effects that can also be sought in the context of economic policy'' (\S61). The PSPP was also held to be proportionate (\S71ff) and respecting the prohibition to grant overdraft facilities or any other type of credit facilities of Article 123 TFEU (\S101ff).

This time, however, the GFCC refused to accept the CJEU’s ruling. In its so-called PSPP judgment, the GFCC found that the Federal Government of Germany and the Bundestag violated the complainants' constitutional rights by failing to take steps challenging the ECB's PSPP decision (GFCC, 2 BvR 859/15, May 5, 2020; PSPP judgement). The GFCC declared both the ECB's decision and the CJEU's judgment upholding the decision as \textit{ultra vires} acts depriving them of their ``binding force in Germany'' (\S\S154, 163 and 165). Such an \textit{ultra vires} act of the ECB or the CJEU is ``not to be applied in Germany, and has no binding effect in relation to German constitutional organs, administrative authorities and courts'' (\S234). The consequences affect, \textit{inter alia}, also the Bundesbank: ``[...] Following a transitional period of no more than three months [...], the Bundesbank may thus no longer participate in the implementation and execution of the PSPP Decision'' (\S235).

The story does not end here. Shortly after, an order of execution was sought with the GFCC. The GFCC found the application inadmissible (GFCC, Order of the Second Senate, 2 BvR 1651/15, 2BvR 2006/15, April 29, 2021). To the surprise of many, the Court explained at length that the applications were not only inadmissible but also unfounded. In essence it explained that the Federal Government and the Bundestag dispose of a wide range of measures to fulfil the requests and that the measures taken did indeed fulfil the PSPP judgement's requirements (\S108). It also found that the ECB Governing Council's assessment of the proportionality of the PSPP had been sufficient (\S100). This can be seen as the GFCC's ``final green light'' for the PSPP. The European Commission, though, was not satisfied. On June 8 2021, the Commission sent a letter of formal notice (under Article 258 TFEU) to Germany for breach of fundamental principles of EU law, because of the GFCC's PSPP judgment. Further developments remain to be seen.

This discussion shows that courts have the potential to severely limit the possibilities and flexibility of the ECB to fight adverse economic or other events. In the next step, we will develop an econometric model with the goal to measure the effects associated with these rulings on a large panel of mostly financial indicators for selected euro area countries.

\section{Measuring central bank independence shocks}\label{sec: econometrics}
Table \ref{tab:rulingdates} provides us with several event dates (``Publication'') which we can use to analyze the effects of ruling announcements on a set of outcome variables. Since the number of events is rather low, reliable estimation of the causal effect of the rulings is difficult and calls for appropriate techniques which we develop in this section. Our econometric framework builds on the notion that rulings introduce uncertainty at high (i.e., daily) frequencies and these increases in volatility (relative to days without significant economic/political events that might impact financial markets) are used to identify the causal effect of legal challenges to the ECB's independence. Our approach fundamentally builds on and extents the model developed in  \citet{gurkaynak2020missing}. 

Let $\bm y_t = (y_{1t},\hdots,y_{Mt})'$ denote a panel of $M$ daily variables which, for simplicity, are assumed to have an unconditional mean equal to zero.  We model $\bm y_t$ using a static factor model with time-varying factor loadings:
\begin{equation}
    \bm y_t = d_t \bm \lambda f_t + \bm \varepsilon_t,\quad \bm \varepsilon_t \sim \mathcal{N}(\bm 0, \bm \Sigma), \label{eq: obs}
\end{equation}
with $d_t\in\{0,1\}$ being an indicator which equals one if $t$ is in a so-called event (or ruling) set $\mathcal{A}_{\text{R}}$ while it equals zero if $t$ is included in a non-event set $\mathcal{A}_{\text{NE}}$. $\bm \lambda$ denotes an $M$-dimensional vector of factor loadings and $\bm{\varepsilon}_t$ is a Gaussian shock vector with zero mean and a diagonal variance-covariance matrix $\bm \Sigma =\text{diag}(\sigma_1^2, \dots, \sigma_M^2)$. $f_t \sim \mathcal{N}(0, 1)$ is a latent factor which summarizes co-movements in the yields on ruling dates. The key implication of Eq. (\ref{eq: obs}) is that on ruling dates, the implied variance-covariance matrix of $\bm y_t$ is:
\begin{equation}
    \text{Var}(\bm y_t) = \bm \lambda \bm \lambda' + \bm \Sigma, \label{eq: commonalities}
\end{equation}
whereas on non-event dates, it is simply $\text{Var}(\bm y_t)=\bm \Sigma$. Hence, if $\bm \lambda \neq \bm 0_M$, our model implies that the variance of the $i$th element of $\bm y_t$, $y_{it}$, on a ruling day is equal to $\lambda^2_i + \sigma_i^2$ and thus exceeds the variance on a non-event day. The off-diagonal elements of $\text{Var}(\bm y_t)$ on a ruling day are given by $\lambda_i \lambda_j$ for $i \neq j$, which results in a full variance-covariance matrix. By contrast, on non-event days, the shocks in $\bm \varepsilon_t$ are independent. In principle, if movements in $\bm y_t$ are dominated by legal rulings on a specific ruling date, the corresponding commonalities given by
\begin{equation}
    \frac{\lambda^2_{i}}{(\lambda_i^2 + \sigma_i^2)}
\end{equation}
approach unity. 

Model estimation can be carried out using maximum likelihood or Bayesian techniques. In this paper, we follow a Bayesian approach and employ a Gibbs sampler to simulate from the joint posterior of $\bm \lambda, f_t$ and $\bm \Sigma$. This algorithm, however, relies on knowing the non-event set $\mathcal{A}_{\text{NE}}$. The next sub-section describes how this set is formed.

\subsection{Identifying non-event windows}
One important assumption underlying our modeling framework is that we assume the variance of $\bm y_t$ to differ between ruling and non-ruling dates. This is a rather strong assumption. Several important regular events (such as press conferences after governing council meetings, see, e.g., \citealp{altavilla2019measuring}, or data releases by statistical agencies) or non-regular events (such as Mario Draghi's ``Whatever it takes'' speech) have pronounced effects on financial variables. Having such observations in the non-event/non-ruling set would certainly lead to a situation where the assumption of a higher variance on ruling dates is difficult to maintain. This is the reason why we meticulously avoided defining non-ruling windows but instead differentiated between ruling and non-event windows. Moreover, if the number of ``event'' days is small to moderate (which is the case for court rulings), simply discriminating between ruling and non-ruling dates easily masks differences in variances. This makes identifying the factor (and/or the loadings) difficult and renders inference imprecise. 

In the related literature it is common to avoid corrupting event windows of interest with other events by specifying tight windows surrounding these announcements \citep[see, e.g.,][]{kuttner2001monetary, gurkaynak2005sensitivity, gertler2015monetary, jarocinski2020deconstructing}. This paper proposes a different route which does not rely on using intra-day data and/or precise timings of the rulings. Let $\mathcal{A}_{\text{NR}}$ denote the set which includes all days except the ones which feature a ruling decision (i.e., for all $t \notin \mathcal{A}_\text{R}$). We assume that $\bm y_t$ for $t \in \mathcal{A}_{\text{NR}}$ evolves according to:
\begin{equation}
    \bm y_t \sim \sum_{j=1}^J w_j \mathcal{N}(\bm y_t|\bm{0}_M, \vartheta_j^2~\bm \Omega)\quad \text{ for } t \in \mathcal{A}_{\text{NR}},\label{eq: mixmodel}
\end{equation}
with $w_j$ denoting a weight with $w_j \ge 0$ and $\sum_{j=1}^{J} w_j = 1$ while $\vartheta_j^2$ denotes a component-specific variance with $\vartheta_1 < \dots < \vartheta_J$ and $\bm \Omega = \text{diag}(\omega_1^2,\hdots,\omega_M^2)$ is a known diagonal matrix which serves to control for differences in the scaling of $\bm y_t$.\footnote{The ordering restriction on the variances solely refers to the labeling of the regimes, and is a technical restriction that does not affect posterior distributions \citep[see][for details]{fruhwirth2006finite}.} This implies that $\bm\Sigma = \vartheta_1\bm{\Omega}$ in terms of Eq. (\ref{eq: obs}), featuring solely the observations in the minimum variance regime, i.e., the set $\mathcal{A}_{\text{NE}}$. In what follows, the diagonal elements of $\bm \Omega$ are set equal to the empirical variances of the different columns in $\bm Y = (\bm y_1, \dots, \bm y_T)'$. 

It is noteworthy that $\vartheta_j$ serves as a common scalar factor which drives the variances on non-ruling dates.  To see this more clearly, Eq. (\ref{eq: mixmodel}) can be equivalently stated in terms of auxiliary variables $\delta_t$ which select the mixture component:
\begin{equation}
    \bm y_t|\delta_t = j \sim \mathcal{N}(\bm y_t| \bm{0}_M, \vartheta^2_j ~\bm \Omega), \label{eq: mixrep}
\end{equation}
which implies that if $\delta_t = j$, the variance of $\bm y_t$ is proportional to $\vartheta_j^2$. Under the ordering restriction that the variances are increasing, the set $\mathcal{A}_{\text{NE}} = \{t: \delta_t=1\}$ collects all time periods where the variance of the process in Eq. (\ref{eq: mixrep}) is minimal. Our key assumption is that on days in the non-event set, no substantial new information is driving movements in the variables collected in $\bm{y}_t$.

This, of course, relies on an appropriate specification of $J$. Too small values of $J$ could lead to regimes that might still be contaminated by other/non-ruling based events (and thus feature a too high variance) whereas too large values of $J$ could translate into regimes with too few observations in the non-event set. In our case, we use a Bayesian approach in the spirit of \citet{malsiner2016model} and estimate $J$ by setting it to a large value (i.e., $J=30$) and then use shrinkage to empty out irrelevant clusters. For the remaining parameters of the model, we choose adequate weakly informative priors. Appendix \ref{app:A:econometrics} provides further details on the prior setup and the Markov chain Monte Carlo sampling algorithm.

\subsection{Local projections to compute dynamic responses}
We use the general model in Eq. (\ref{eq: obs}) specified as a local projection \citep[see][]{jorda2005estimation} for horizons $h=0,1,\hdots,H$, such that we can interpret the factor loadings in $\bm{\lambda}_h=(\lambda_{1h},\hdots,\lambda_{Mh})'$ directly as the impulse response function for variable $y_{jt}$ at $h$. Note that our variable set is comprised mostly of financial variables. We thus assume that their dynamic evolution is described well with a random walk, and consequently consider the dependent variable in differences, $\Delta y_{j,t+h}$, similar to, e.g., \citet{lewis2019monetary,swanson2021measuring}:\footnote{Experiments indicate that using this type of linear projection rather than including a set of (lagged) control variables produces very similar results, pointing to the adequacy of the random walk assumption.}
\begin{equation*}
    (y_{j,t+h} - y_{j,t-1}) \equiv \Delta y_{j,t+h} = d_t\lambda_{jh} f_{t+h} + \epsilon_{j,t+h}.
\end{equation*}
We consider a horizon of up to $H=60$ (corresponding to three months of workday-weeks). {Specifying the model directly as a local projection has the advantage of taking all sources of uncertainty into account within a unified econometric framework. By contrast, a two-stage approach, estimating the factor first and using, e.g., its posterior median to compute dynamic responses with local projections, may yield incorrect inference and too narrow credible sets for parameter estimates.} 

\section{The effects of central bank independence shocks}\label{sec: emp_results}
In this section we present our empirical findings. The first sub-section briefly discusses the dataset while Section \ref{sec: model_features} presents features of our model such as the estimated latent factor, the cluster allocation and the idiosyncratic variances. Sub-section \ref{sec: IRFs} presents the dynamic responses to central bank independence shocks {(we also refer to these interchangeably as ruling shocks).}

\subsection{Data overview}
To understand the impact of the central bank independence shocks we study a representative set of European, US and international financial variables observed at a daily frequency. Our sample covers the period from March 2012 to March 2021.

Among other variables we consider in our discussion of dynamic responses to ruling shocks is the CISS, which is the Composite Indicator of Systemic Stress in the financial system \citep[see][]{hollo2012ciss}. It is a summary measure aggregating indicators of financial stress in the banking sector, money, equity, bond and foreign exchange markets. To measure sentiments in the sovereign debt markets we include the yields on 10-year government bond yields of the euro area member states. The risk-free yield curve, based on the overnight index swap (OIS) rates at maturities from 1 month to 2 years reflects the market expectations regarding ECB policies. European inflation swaps at maturities 1 to 10 years combine market expectations of inflation and inflation risk premia. European stock market developments are represented by the EuroStoxx 50 index accompanied by headline country-indices. Corporate bond market developments are reflected by the Option-Adjusted Spreads (OAS) of high-yield, BBB and AA corporate bonds over the sovereign bonds of the same maturity, constructed by ICE BofAML. 

{Moreover, the information set also includes several other non-domestic and international variables such as gold and oil prices, or financial variables for the US. A full list of variables and additional information are provided in Appendix \ref{app:B:data}. Series and country-coverage is governed by data availability on a daily frequency.}

\subsection{Ruling dates from an econometric perspective}\label{sec: model_features}
Our model yields a posterior distribution over the factors $f_t$ for different values of $h$. These can be interpreted as an unexpected shock to central bank independence induced by legal events. Since the factor is not identified up to the sign, economic interpretation needs to be carried out in light of the estimated factor loadings in $\bm \lambda_h$. This is what we do in Section \ref{sec: IRFs}. In this section, we focus on the quantitative properties of the factor and assess whether using the mixture model sharpens statistical inference.
\begin{figure}[t]
    \includegraphics[width=\textwidth]{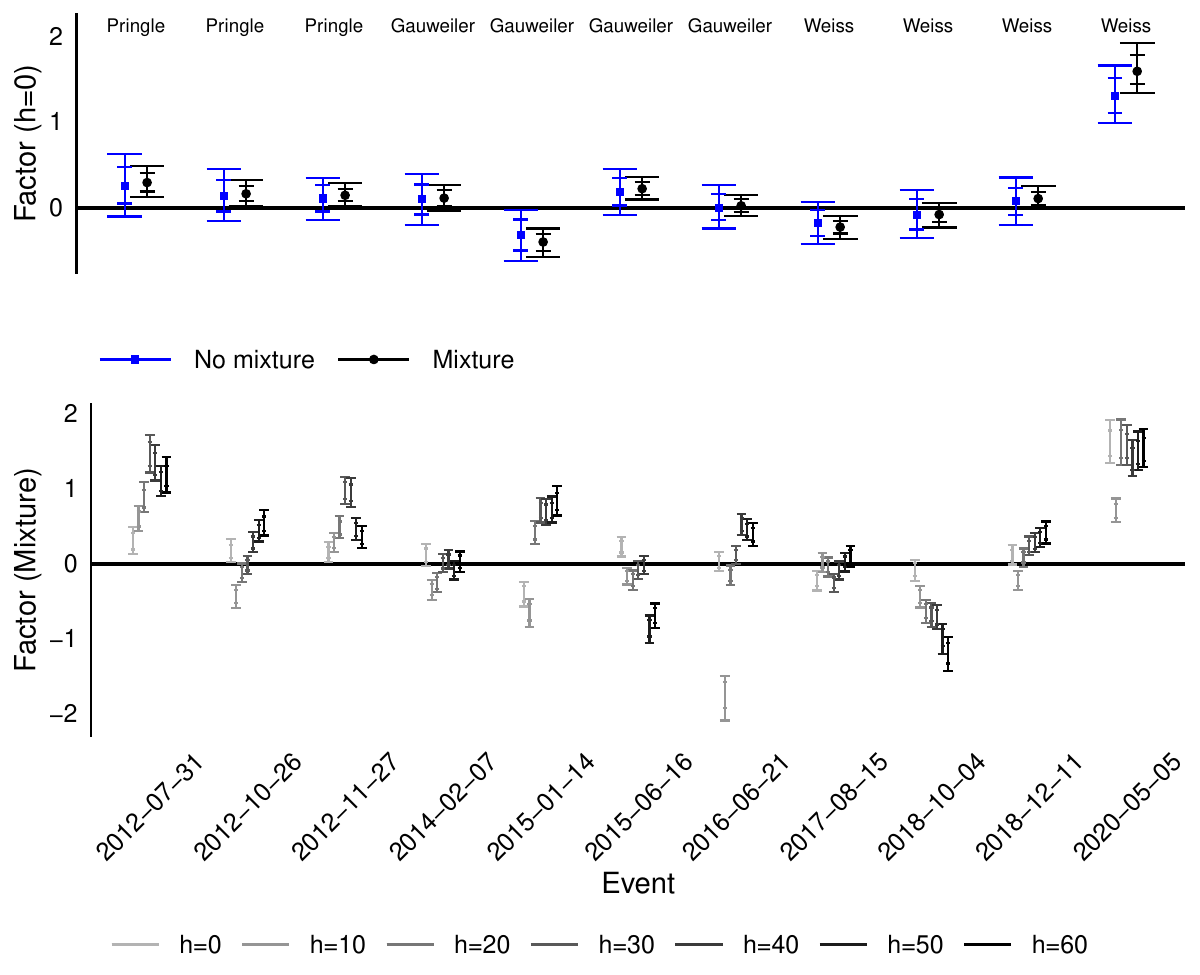}
    \caption{Estimates of the factor $f_{t+h}$ at ruling dates. Upper panel: The black and blue errorbars (horizontal lines mark 68 and 90 percent posterior credible sets) are shown alongside the posterior median for a non-mixture and mixture specification with $J=30$. Lower panel: Horizon-specific 68 and 90 percent posterior credible sets of the factor.}
    \label{fig:factor}
\end{figure}

Figure \ref{fig:factor} depicts boxplots of the posterior distribution of $f_t$ for $h=0, 1, \dots, H$ across the different events we use to identify the shock. To investigate the effect the mixture model has on the estimated factors, we include boxplots of a model that naively discriminates between ruling and non-ruling days (in black) and compare them to our approach which uses a mixture model to group non-event days into a single cluster (in blue). The figure suggests that if we do not purge non-ruling days from other events, the corresponding uncertainty bounds are much wider and include zero in almost all cases. Two cases in which zero is not included are January 14, 2015 (Gauweiler) and May 5, 2020 (Weiss). In the latter case, the GFCC refused to accept the CJEU's ruling and challenged the PSPP of the ECB. 

If we estimate $f_t$ based on the model discussed in Section \ref{sec: econometrics} we find that posterior uncertainty decreases appreciably in all cases considered. Comparing the sign of the posterior medians reveals that both approaches yield similar findings in terms of the direction of the shock but, in terms of magnitudes, we find the estimates based on the mixture model being slightly more pronounced.  Again, the strongest reaction of the factor is observed on May 5, 2020.

\begin{figure}[t]
    \includegraphics[width=\textwidth]{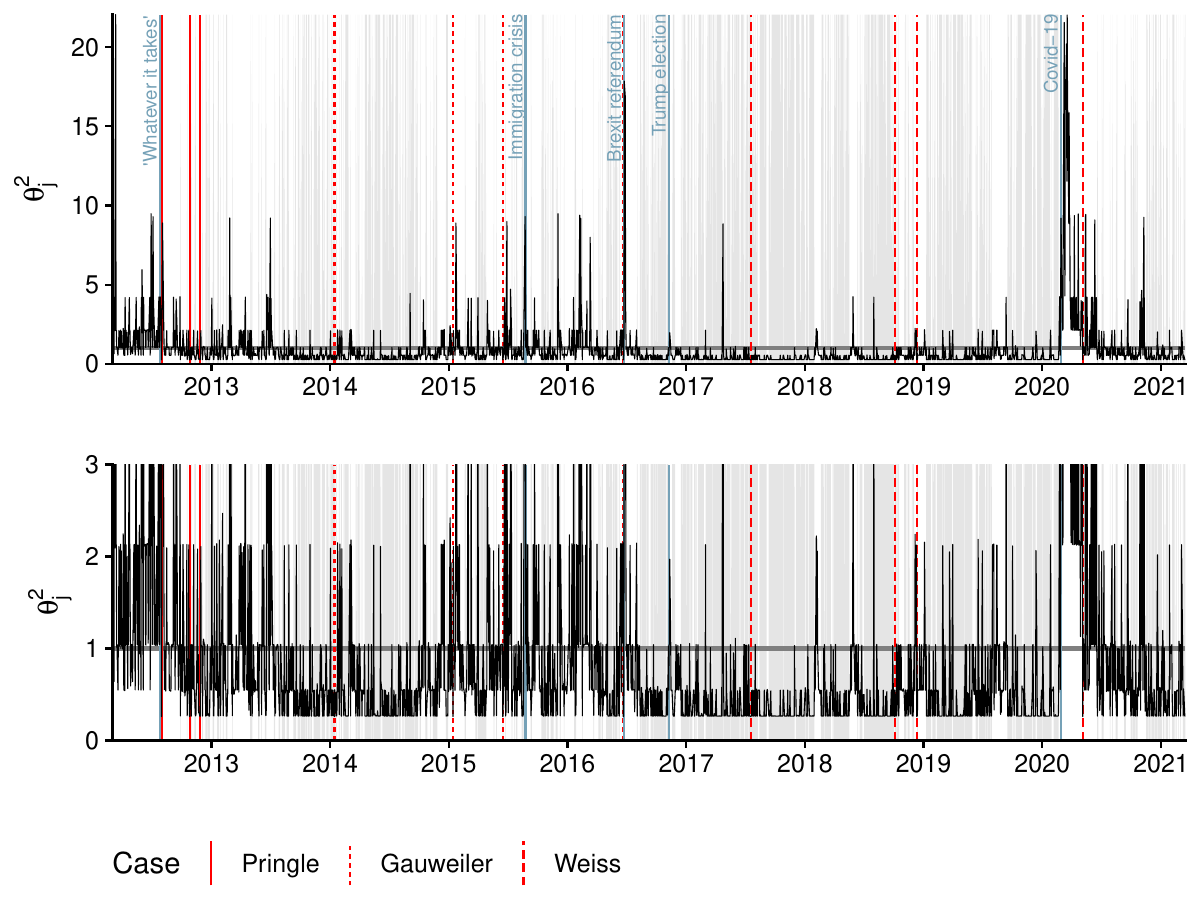}
    \caption{Variance clusters and the non-event set alongside ruling dates for $h=0$. The solid black line plots $\theta_j^2$ over time (the solid dark grey horizontal line marks unit variance). The grey shaded area indicates the posterior median of a specific period being featured in the minimal-variance regime (and thus, the non-event set). The red vertical lines are the ruling dates from Table \ref{tab:rulingdates}, the light blue vertical lines are other known events that moved financial markets.}
    \label{fig:minvar}
\end{figure}

Next we consider the estimated variance clusters as well as the non-event set. These are contrasted with the ruling dates and shown in Figure \ref{fig:minvar} for $h=0$. The gray shaded areas in the figure refer to the probability that a specific period is included in the non-event set. The red lines refer to the different ruling dates while the light blue lines mark important events that moved financial markets.  The upper panel shows the full range of the estimated volatilities $\theta^2_j$ over time (i.e., for the time $t$ variance if $\delta_t=j$) whereas the lower panel restricts the range of the graph. For robustness checks varying the number of mixture components, see Appendix \ref{app:C:results}.

From this figure, it is clearly visible that contrasting ruling dates with non-ruling dates is insufficient. In this case, other important events and volatility spikes mask potential effects and  lead to imprecise inference on the effect of the ruling dates. In principle, we find that overall volatility spikes in the case of several important events and, if these are to dominate the estimates of the idiosyncratic volatilities, the explanatory power of the ruling factor is greatly diminished. By contrast, the common idiosyncratic variance estimated from including only information included in the non-event  and ruling sets is small and the corresponding probabilities that a given day is located in this minimum variance-set is, in most cases, close to one.

\begin{figure}[!t]
    \includegraphics[width=\textwidth]{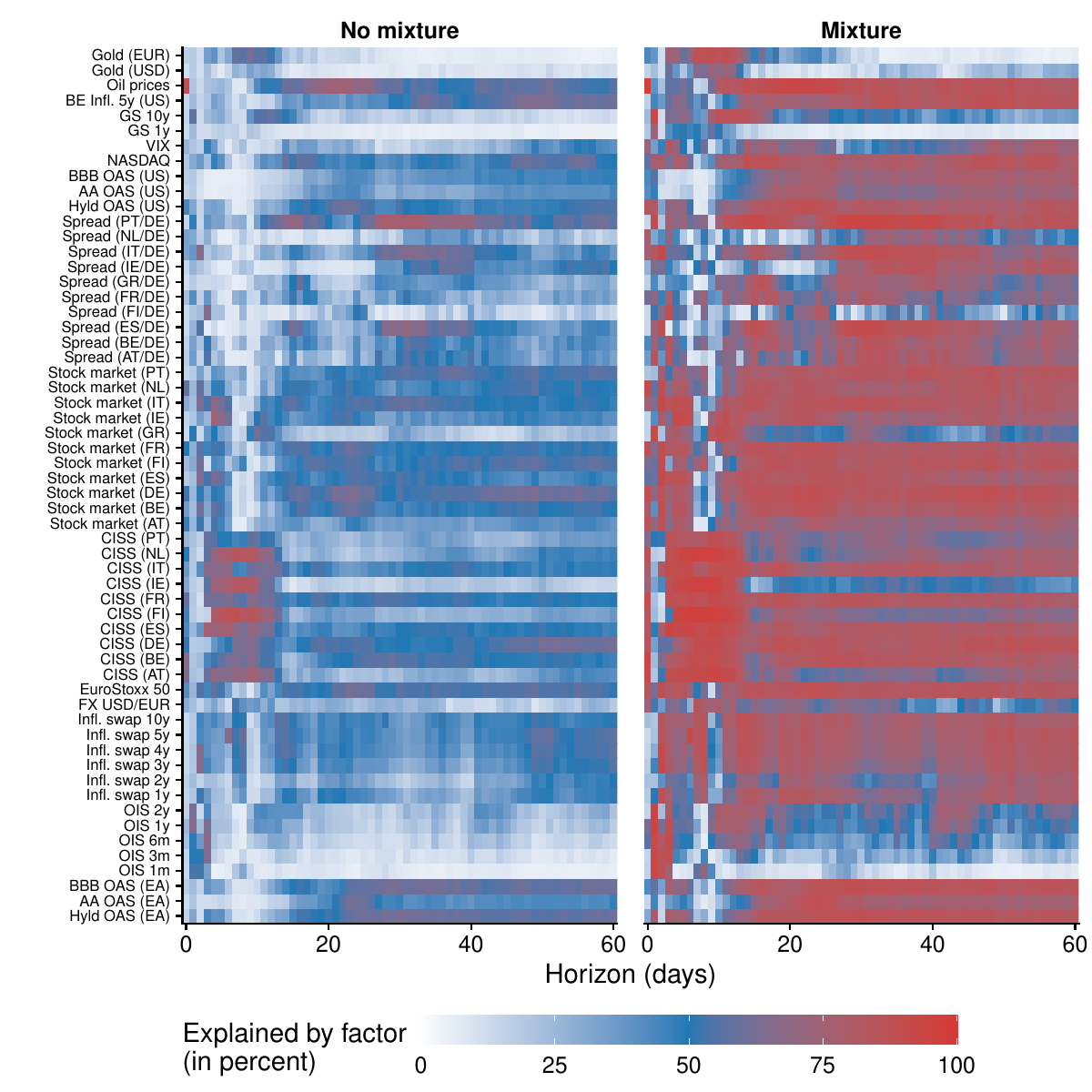}
    \caption{Share of the variance explained by the factor, $\lambda_{jh}^2/(\lambda_{jh}^2 + \sigma_{jh}^2)$, at event dates for the model with and without the mixture specification.}
    \label{fig:facexpl}
\end{figure}

If our approach is successful, the commonalities in Eq. (\ref{eq: commonalities}) should be sizable (i.e., close to unity). By contrast, if we do not systematically control for other events on non-ruling days, the estimate of $\sigma_i^2$ will be large and $\lambda_i$ relatively small. We investigate this in Figure \ref{fig:facexpl}. The figure  depicts the share of variance explained through the common factors on ruling dates for a model without the mixture setup and one that searches for a minimum variance regime. These quantities are computed for each forecast horizon white cells indicate commonalities close to zero whereas red cells suggest that commonalities approach one.

Considering the left panel of Figure \ref{fig:facexpl} reveals a great deal of white and blue colored cells. For some quantities (e.g., gold, GS1, OIS 1m) we find that the factor explains little variation on ruling days (especially for longer forecast horizons) but a sizable amount of variation is explained in the short term (with values close to 50\%).  The opposite is true for several other variables such as stock markets, spreads \textit{vis-\'{a}-vis} German treasuries, as well as inflation swaps across different maturities. For these quantities, we find that the factor explains around 50\% of total variation on ruling days after around 10 to 20 days. Interestingly, when we consider the reactions of country-specific financial conditions (measured through the CISS), commonalities reach values close to one for horizons between five to 10 days. For short-run reactions and longer-run reactions, these shares are considerably lower. This indicates that, even without controlling for the fact that the non-ruling set is corrupted by several variance outliers, we find sizable shares explained through the ruling factor. The fact that CISS reactions are mostly dominated by ruling shares after a few days implies that for these horizons, news about rulings dominate movements in financial conditions but this effect quickly dies out afterwards.

We now turn to the question on how the mixture model impacts commonalities. The right panel of Figure \ref{fig:facexpl} suggests that for most quantities, we observe variance shares explained which are close to one. At a general level, we find that immediate reactions often feature shares which are appreciably below one but for higher order responses the pattern is quite consistent across variable-types and countries (with some exceptions being gold in EUR and USD, GS 1y and 10y, the yield spread between Finland and Germany and the CISS for Ireland). For most of these it makes sense that the corresponding factor loadings are close to zero and hence the factor plays a limited role in explaining time series variation. 

Summing up this discussion, we find that even without using the mixture model some variables display sizable commonalities for several impulse response horizons. In principle, adding the mixture model and thus purging the non-event set from other, non-ruling related events which had a pronounced impact on financial markets, further increases the explanatory power of our approach.

\subsection{Impulse response functions from local projections} \label{sec: IRFs}
\begin{figure}[t]
    \includegraphics[width=0.24\textwidth]{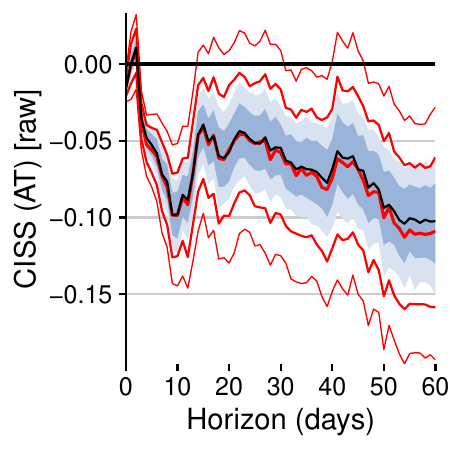}
    \includegraphics[width=0.24\textwidth]{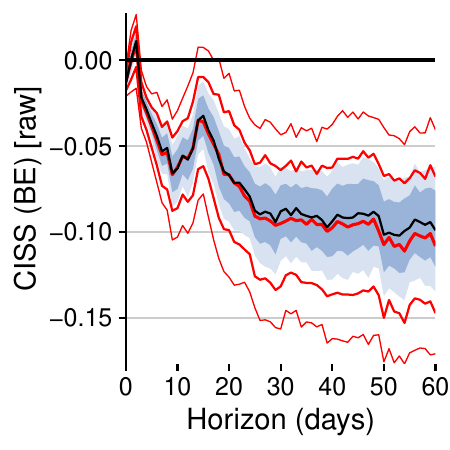}
    \includegraphics[width=0.24\textwidth]{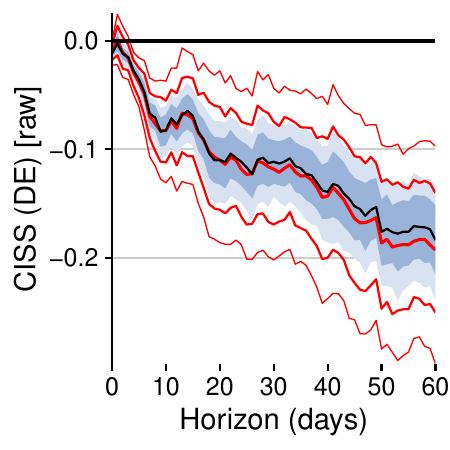}
    \includegraphics[width=0.24\textwidth]{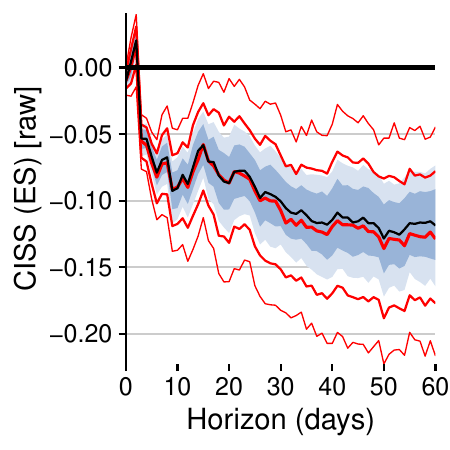}
    \includegraphics[width=0.24\textwidth]{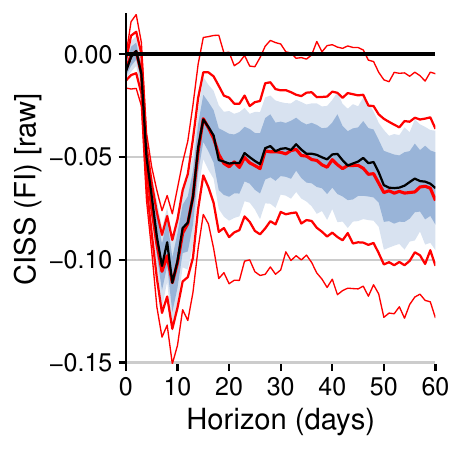}
    \includegraphics[width=0.24\textwidth]{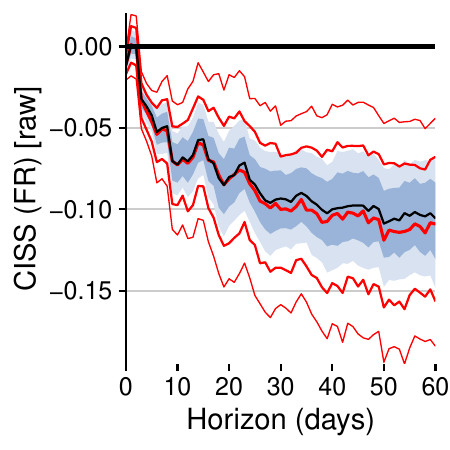}
    \includegraphics[width=0.24\textwidth]{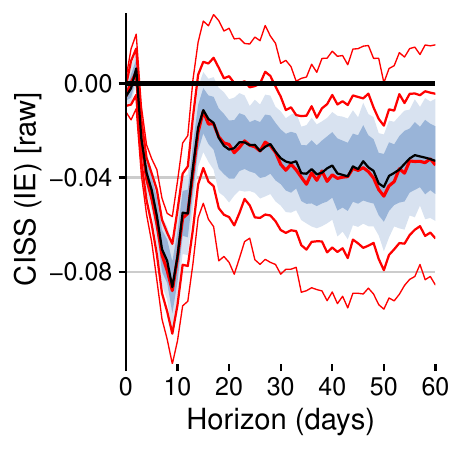}
    \includegraphics[width=0.24\textwidth]{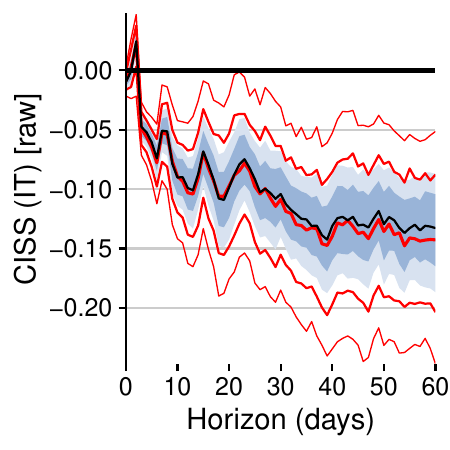}
    \includegraphics[width=0.24\textwidth]{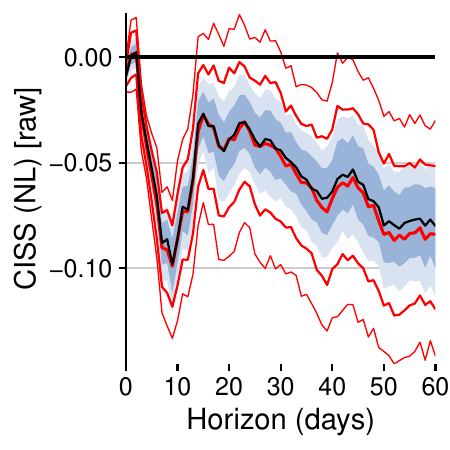}
    \includegraphics[width=0.24\textwidth]{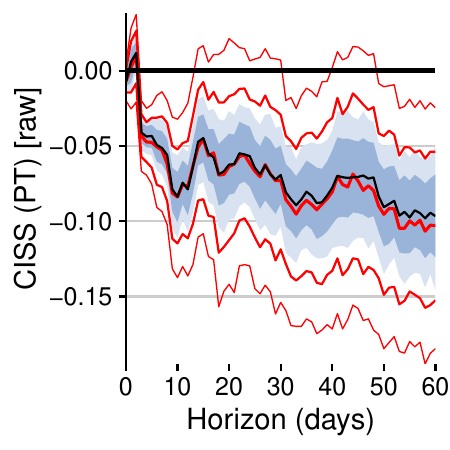}
    \caption{Impulse responses for CISS measures across euro area economies. The solid black line shows the posterior median estimate for the mixture model, alongside the 68 and 90 percent posterior credible sets in shades of blue. The red lines mark the posterior median, $16$/$84$th and $5$/$95$th percentile for the non-mixture specification.}
    \label{fig:irf-CISS}
\end{figure}

\begin{figure}[t]
    \includegraphics[width=0.24\textwidth]{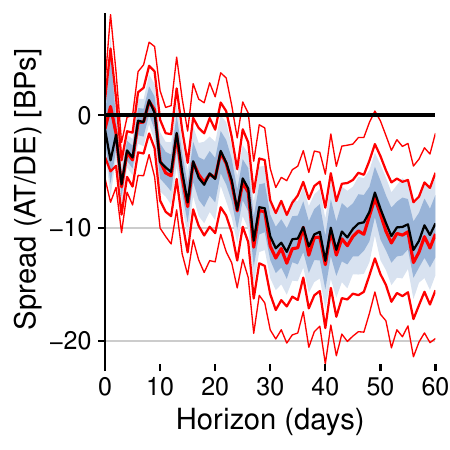}
    \includegraphics[width=0.24\textwidth]{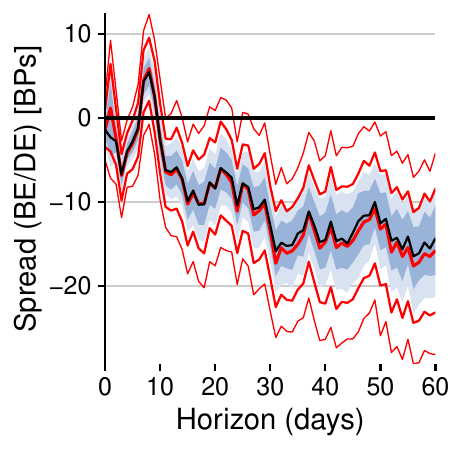}
    \includegraphics[width=0.24\textwidth]{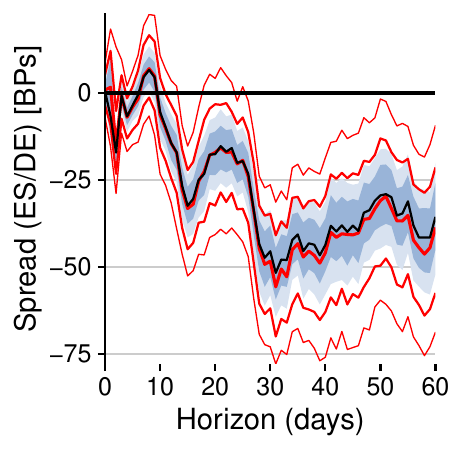}
    \includegraphics[width=0.24\textwidth]{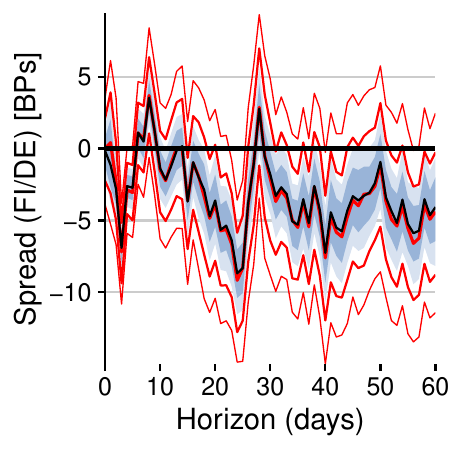}
    \includegraphics[width=0.24\textwidth]{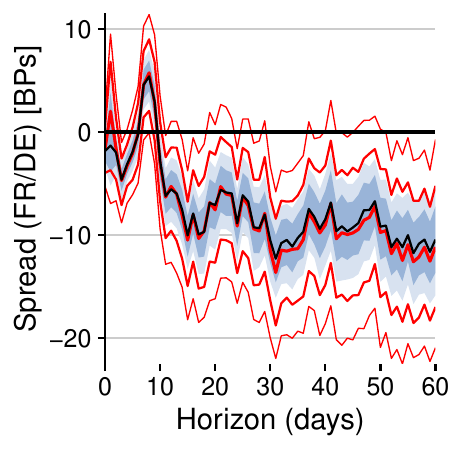}
    \includegraphics[width=0.24\textwidth]{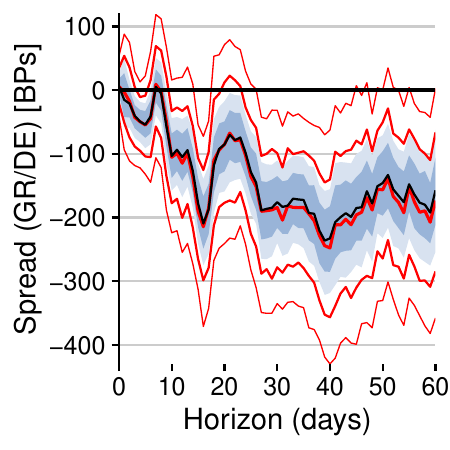}
    \includegraphics[width=0.24\textwidth]{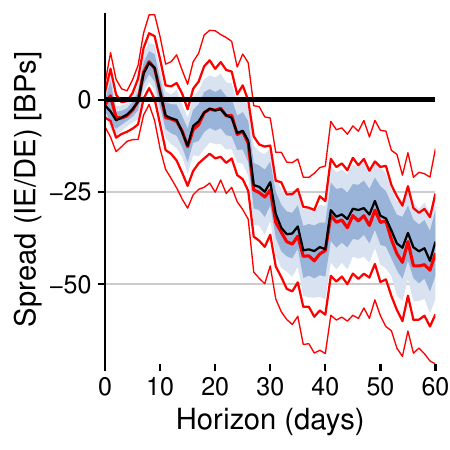}
    \includegraphics[width=0.24\textwidth]{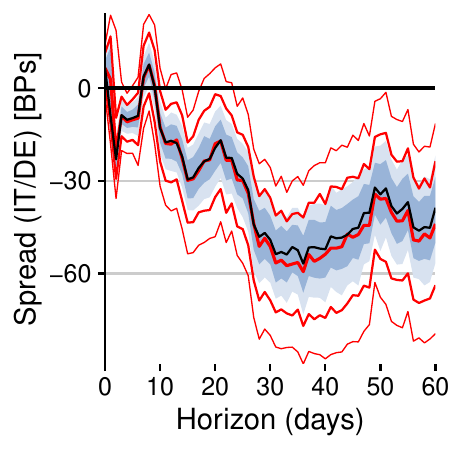}
    \includegraphics[width=0.24\textwidth]{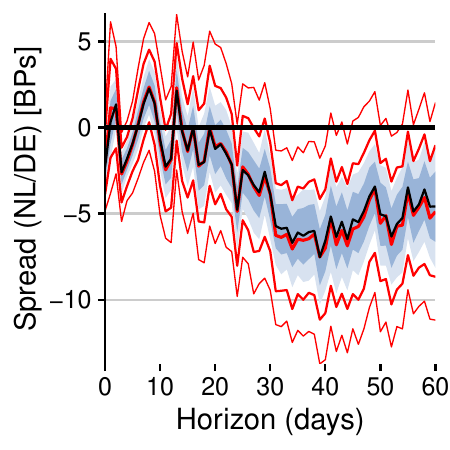}
    \includegraphics[width=0.24\textwidth]{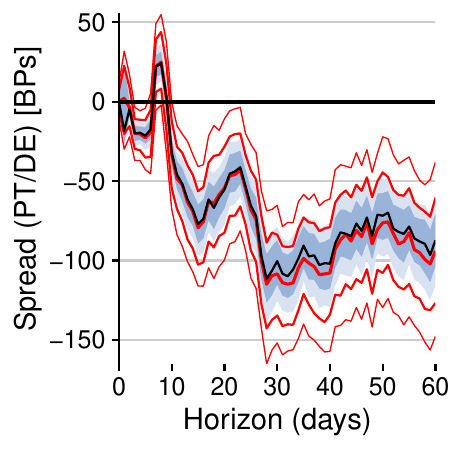}
    \caption{Impulse responses for spreads of domestic government bond yields relative to Germany. The solid black line shows the posterior median estimate for the mixture model, alongside the 68 and 90 percent posterior credible sets in shades of blue. The red lines mark the posterior median, $16$/$84$th and $5$/$95$th percentile for the non-mixture specification.}
    \label{fig:irf-spreads}
\end{figure}

\begin{figure}[t]
    \includegraphics[width=0.24\textwidth]{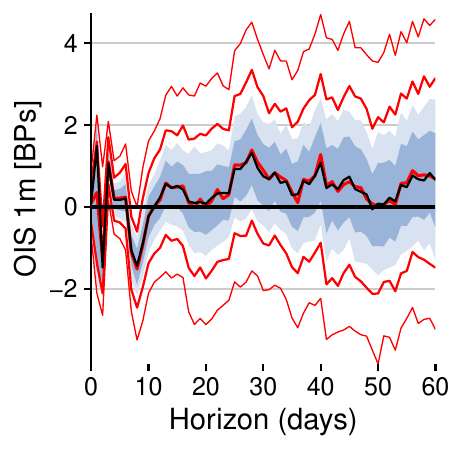}
    \includegraphics[width=0.24\textwidth]{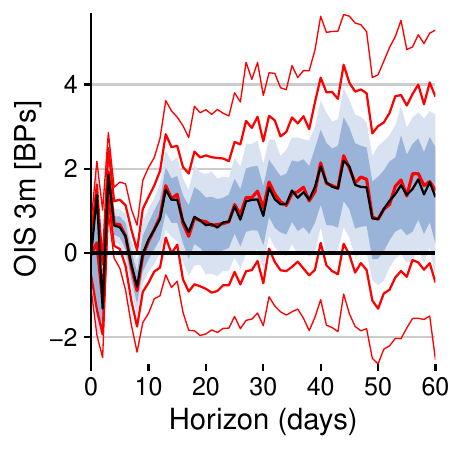}
    \includegraphics[width=0.24\textwidth]{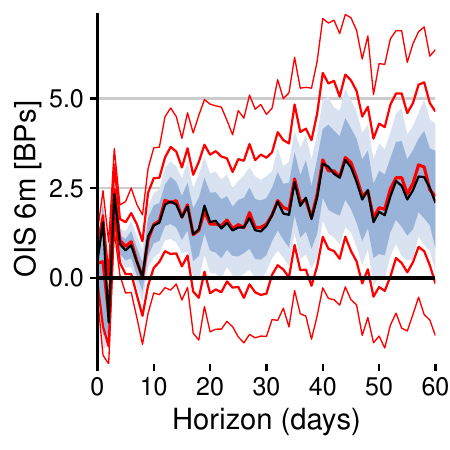}
    \includegraphics[width=0.24\textwidth]{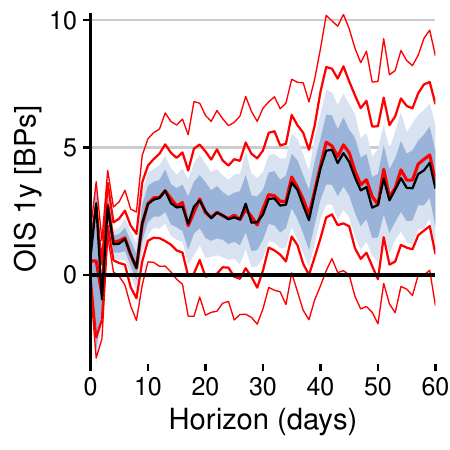}
    \includegraphics[width=0.24\textwidth]{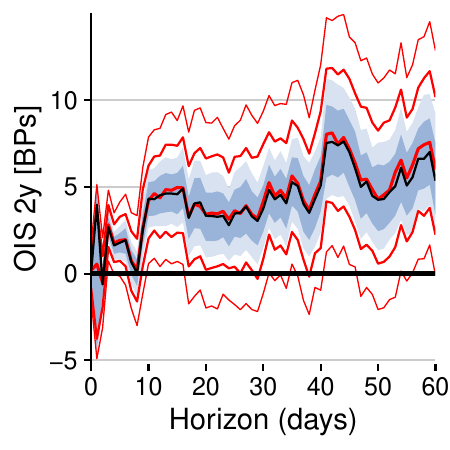}
    \includegraphics[width=0.24\textwidth]{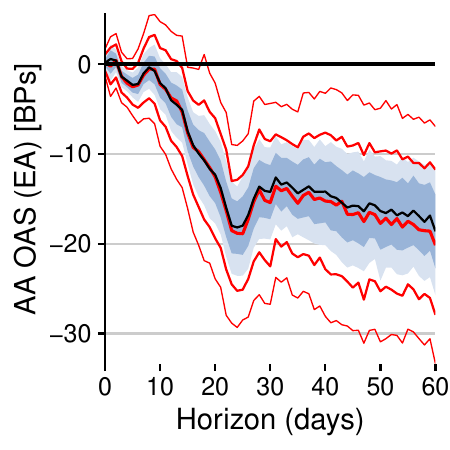}
    \includegraphics[width=0.24\textwidth]{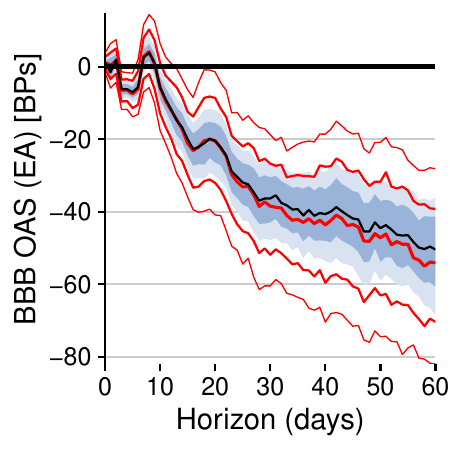}
    \includegraphics[width=0.24\textwidth]{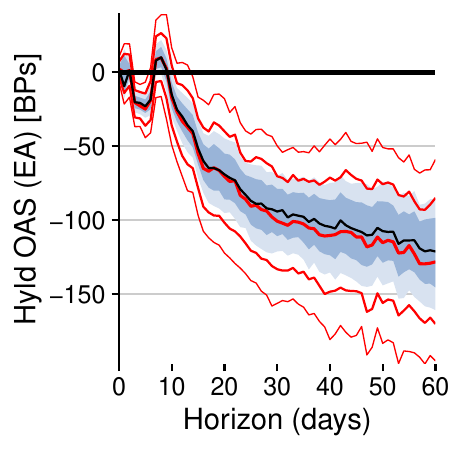}
    \caption{Impulse responses for short-term interest rates and variables measuring financial conditions. The solid black line shows the posterior median estimate for the mixture model, alongside the 68 and 90 percent posterior credible sets in shades of blue. The red lines mark the posterior median, $16$/$84$th and $5$/$95$th percentile for the non-mixture specification.}
    \label{fig:irf-fincond}
\end{figure}

\begin{figure}[t]
    \includegraphics[width=0.24\textwidth]{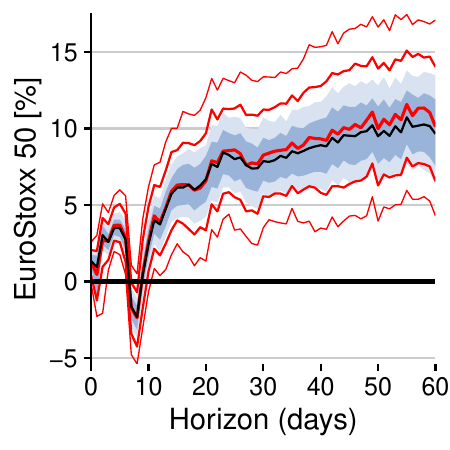}
    \includegraphics[width=0.24\textwidth]{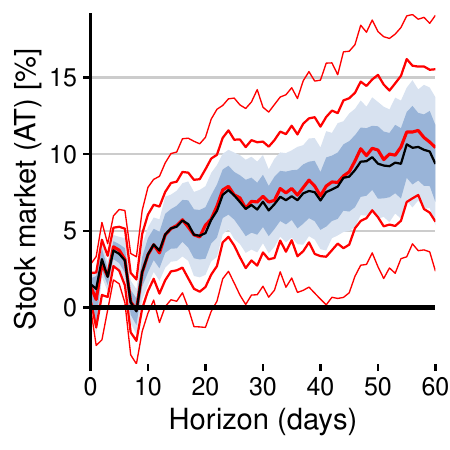}
    \includegraphics[width=0.24\textwidth]{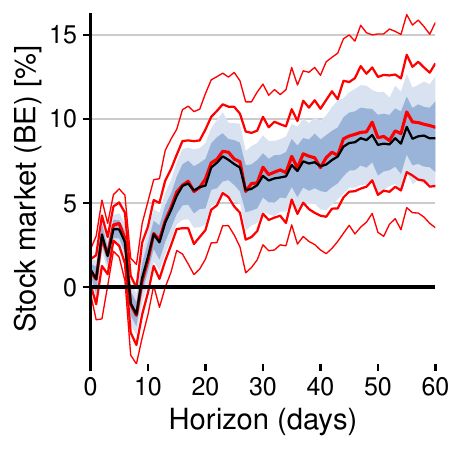}
    \includegraphics[width=0.24\textwidth]{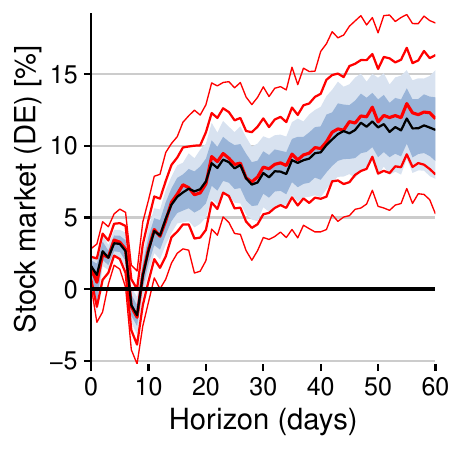}
    \includegraphics[width=0.24\textwidth]{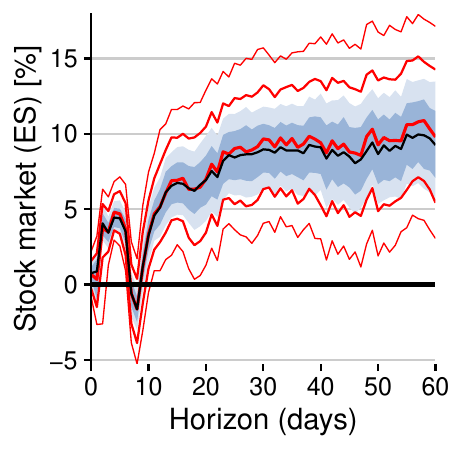}
    \includegraphics[width=0.24\textwidth]{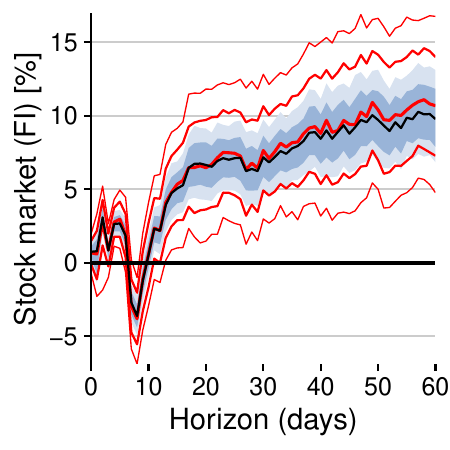}
    \includegraphics[width=0.24\textwidth]{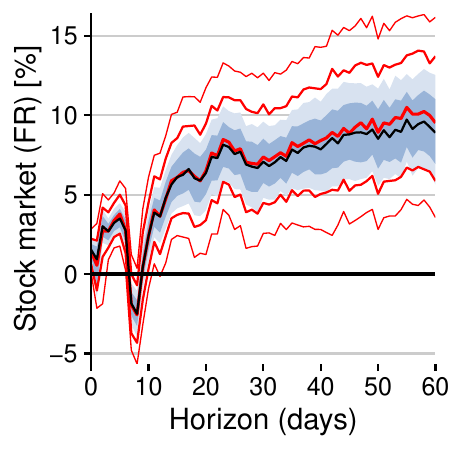}
    \includegraphics[width=0.24\textwidth]{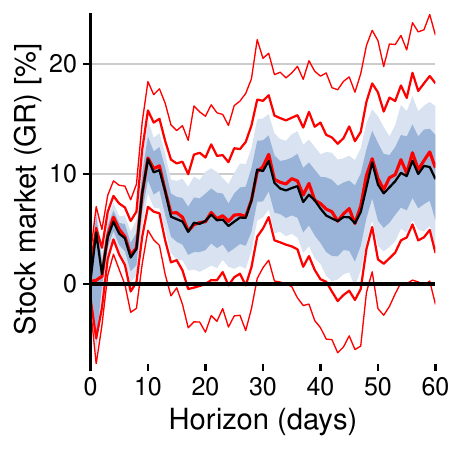}
    \includegraphics[width=0.24\textwidth]{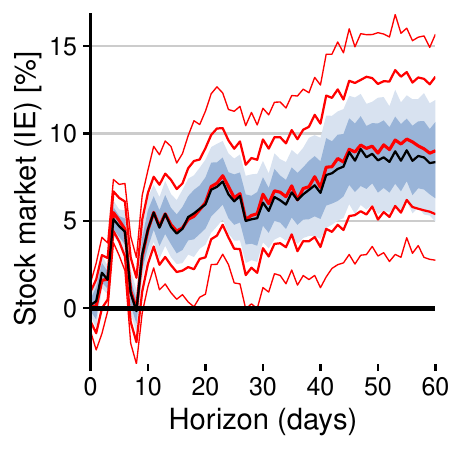}
    \includegraphics[width=0.24\textwidth]{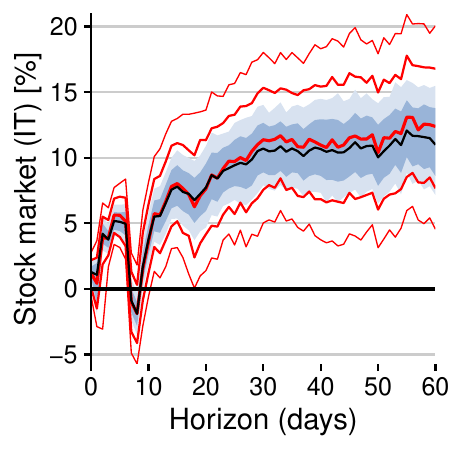}
    \includegraphics[width=0.24\textwidth]{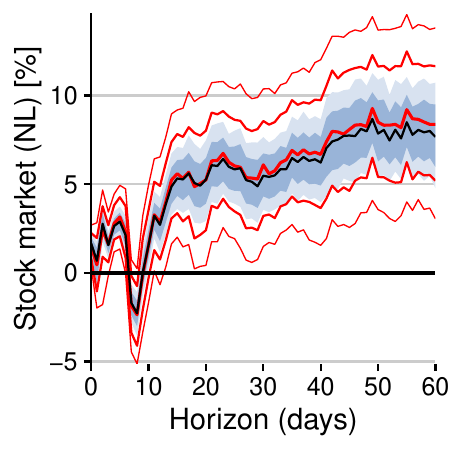}
    \includegraphics[width=0.24\textwidth]{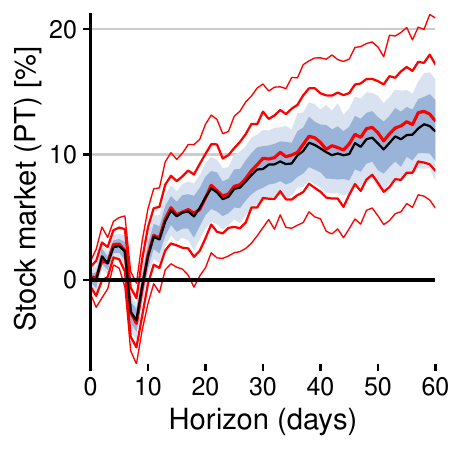}
    \caption{Impulse responses for major stock market indices across euro area economies. The solid black line shows the posterior median estimate for the mixture model, alongside the 68 and 90 percent posterior credible sets in shades of blue. The red lines mark the posterior median, $16$/$84$th and $5$/$95$th percentile for the non-mixture specification.}
    \label{fig:irf-stocks}
\end{figure}

\begin{figure}[t]
    \includegraphics[width=0.24\textwidth]{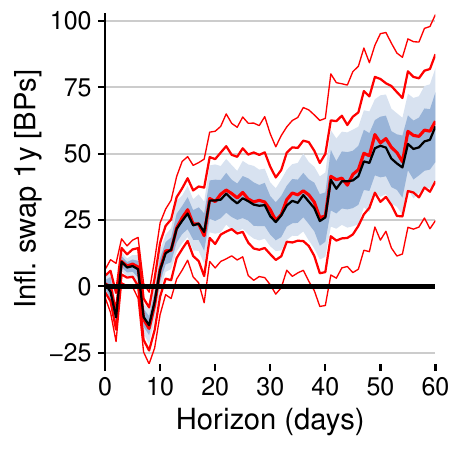}
    \includegraphics[width=0.24\textwidth]{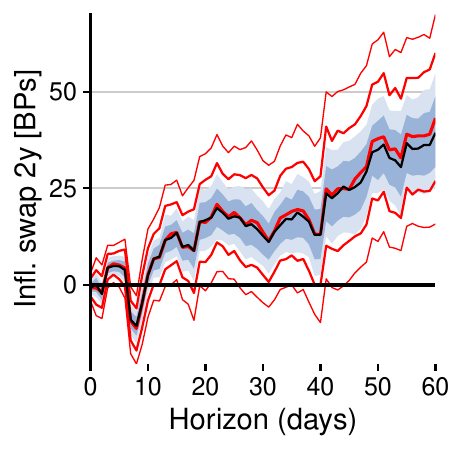}
    \includegraphics[width=0.24\textwidth]{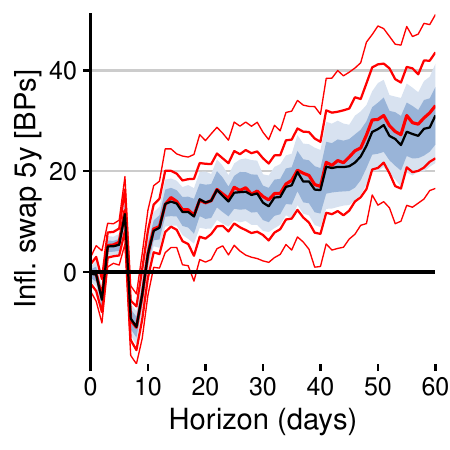}
    \includegraphics[width=0.24\textwidth]{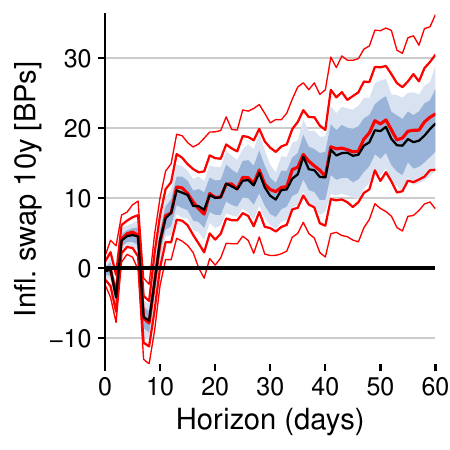}
    \caption{Impulse responses for inflation swaps. The solid black line shows the posterior median estimate for the mixture model, alongside the 68 and 90 percent posterior credible sets in shades of blue. The red lines mark the posterior median, $16$/$84$th and $5$/$95$th percentile for the non-mixture specification.}
    \label{fig:irf-ISyears}
\end{figure}

In this section we discuss the impulse responses to an expansionary ruling shock. With expansionary we mean a situation in which the courts rejected the challenges to the ECB's independence. We normalize the shock such that an expansionary ruling surprise decreases the average cross-country CISS by one standard deviation after 10 days.\footnote{{The majority of rulings in our sample (as opposed to intermediate legal procedures) is in favor of policies enacted by the ECB, and thus, expansionary in nature. This is why we choose, for ease of interpretation, to normalize the shock producing our impulse response functions to be expansionary. It is worth noting that responses in our model framework are symmetric with respect to the sign and size of the shock.}} Figures \ref{fig:irf-CISS} to \ref{fig:irf-ISyears} present the responses. The blue shaded areas represent the 68 and 90 percent posterior credible sets and the black lines the posterior median estimates for up to 60 days after a ruling shock has happened. The red lines depict the posterior quantiles of the model without the mixture specification.

Our prior intuition suggests that an expansionary ruling shock (i.e., increased central bank independence) should decrease financial stress. This is because a decreased likelihood that the ECB is forced to stop its unconventional monetary policy programs translates into a declining probability of sharply increasing spreads between core and peripheral bonds, and provides increased credibility to its accommodative policy stance \textit{vis-\'{a}-vis} financial market participants, affecting their expectations. Consistent with our normalization assumption of the shock and our prior reasoning, Figure \ref{fig:irf-CISS} reports that financial stress as measured by the CISS decreases. When we focus on the shapes and magnitudes of the responses we observe many similarities across the majority of countries: immediate effects appear to be insignificant and after a few days, the CISS starts to decline in a persistent manner, leveling out at around $-0.1$ points after 60 days. The main exceptions from this general pattern are Finland, Ireland and the Netherlands. In these countries, we find that the peak happens much earlier (after around 10 days) and the CISS increases between 10 and 30 days before declining once more.

{The response of the CISS across countries provides valuable insights on the average effects of legal rulings on financial markets in its own right. But these findings are also relevant in terms of the real economy, due to the tight link between financial conditions and economic growth. In particular, the declines we measure in response to an expansionary ruling shock relate to reduced macroeconomic downside-risk and increases in real economic activity \citep[see][]{hollo2012ciss,adrian2019vulnerable,figueres2020vulnerable}.}

Closely related to movements in financial conditions are spreads between 10-year government bond yields of selected euro area member states and Germany, {with the German yields serving as the euro area's risk-free rate. In fact, spreads between core and periphery countries in the euro area are a crucial aspect to be considered by ECB policy makers. This is due to their impact on the smoothness of the transmission of enacted policies to the member states, and they have been among the concerns underlying Mario Draghi's ``Whatever it takes'' speech.} The corresponding responses are depicted in Figure \ref{fig:irf-spreads}. 

Apart from the result that spreads \textit{vis-\'{a}-vis} German treasuries decline for all countries considered, the single most striking observation is the pronounced heterogeneity with respect to the magnitudes of the responses. While core countries such as France, Austria, Belgium or Finland exhibit only modest decreases in spreads which range from 5 to around 15 basis points (BPs), spread reactions in the peripheral countries are enormous. For instance, in the case of Greece we observe that spreads, in response to an expansionary ruling shock, decline by around 250 BPs after around 40 days (with appreciably uncertainty surrounding the point estimates). Greece marks the lower bound. Other countries in that group (such as Italy and Portugal) display substantially smaller reactions which are, however, still significantly larger than the ones observed for the core economies. For Portugal, spreads decline by around 100 BPs after 30 days while Italian spreads decrease by around 60 BPs after around a month. This discussion shows that if the courts reject lawsuits targeted towards stopping QE, on the one hand, financial market participants tend to assess Greek, Italian and Portuguese government bonds more favorably, driving down the yields; on the other hand, financial markets exhibit less demand for German government bond yields. 

The previous discussion has shown that expansionary ruling shocks exert a powerful effect on the longer end of the yield curve. In Figure \ref{fig:irf-fincond} we focus on the reactions at the short-end of the OIS rates with maturities up to 2 years as well as corporate bond spread reactions. Turning to the OIS responses reveals only modest reactions for maturities up to 6 months. They become gradually more pronounced for maturities up to 2 years but with peak reactions of about 5 BPs, the economic significance remains rather low. These subdued reactions can be explained by the fact that during our estimation period, interest rates at the short end of the yield curve have been close to zero and thus display relatively little variation. 

Focusing on the reactions of corporate bond markets across different rating categories (investment-grade and sub-investment-grade) reveals appreciable effects of expansionary ruling shocks which increase with decreasing levels of credit quality. For AA-rated corporate bond spreads the model produces a decline of around 20 BPs after around 60 days. When we consider BBB-rated bonds, these effects double in magnitude, peaking at around 40 BPs. Considering responses of sub-investment grade (high-yield) bond spreads suggests that these decrease in an even more pronounced manner, with peak declines of over 130 BPs. {This relates to our previous discussion of the CISS responses and provides a more nuanced picture of the reactions of credit conditions to ruling shocks.}

Next we consider the reactions of stock markets to a ruling shock. Our intuition suggests that if the ECB is allowed to continue with its unconventional monetary policy programs, stock markets should profit through  different channels. The asset price/wealth channel \citep[see, e.g., ][]{bernanke2005explains, bjornland2009identifying}, the risk taking channel \citep[see, e.g., ][]{borio2012capital} and a general decline in the probability that the ECB has to suddenly stop its unconventional monetary policy operations, with respect to the economic outlook.

When we focus on Figure \ref{fig:irf-stocks} we find empirical evidence which supports our hypothesis that expansionary ruling shocks should benefit stock markets. Across all countries (and for the EuroStoxx 50) we find that stock prices increase (and persistently do so). The general pattern is extremely similar across countries: after no substantial short-run reactions equities increase from day 10 after the shock hit the system. Towards the end of the impulse response horizon the median increases reach about 10 percent. Notice that peripheral countries such as Italy, Portugal and Greece display responses which are observed by slightly more noise. This is driven by the generally higher volatility of the corresponding time series. These stock market reactions to ruling shocks are in line with the findings in \cite{rigobon2004impact} or \citet{bernanke2005explains} who document substantial effects of monetary events on asset prices. 

Finally, all results discussed so far deal with reactions of financial markets. In principle, rejected lawsuits about unconventional monetary policies should also have powerful effects on inflation and inflation expectations. Our working hypothesis is that expansionary ruling shocks, by not constraining the ECB's policy options, should have inflationary effects -- {particularly in light of the unique challenges to monetary policy in the euro area, and our corresponding sampling period which mainly features inflation rates below the ECB's target}.\footnote{{See also the discussion of the impact of central bank independence on inflation dynamics in \citet{crowe2007evolution}. While earlier papers such as \citet{alesina1993central} or \citet{cukierman1992measuring} suggest an inverse relationship between inflation and central bank independence in industrial economies, institutional changes and subdued global inflationary pressures in recent years appear to have reversed this relationship for the case of the euro area.}} This conjecture is confirmed when we consider Figure \ref{fig:irf-ISyears}. In this figure, we show the responses of the Euro inflation swap curve at different maturities. In all these cases, the model produces statistically and economically significant reactions. In terms of magnitudes, we find that short-run inflation expectations increase stronger than longer-run expectations, with increases of about 60 BPs for 1 year and 20 BPs for 10 year swaps.

{Summarising, we find that an expansionary ruling shock eases financial conditions, lowers spreads between core and periphery euro area member states, yields modest increases in short-term rates and inflation expectations, and substantial gains on stock markets. These findings are somewhat reminiscent of responses to contractionary central bank information shocks \citep[see, e.g.,][]{nakamura2018high,jarocinski2020deconstructing}. In particular, our findings relate to ``Delphic'' monetary policy shocks as described in \citet{andrade2021delphic}, which correspond to monetary policy announcements that convey news about future macroeconomic conditions. Given our research design, this points towards financial market participants categorizing expansionary ruling shocks as being beneficial to real activity and the economic outlook.}

% \begin{figure}
%     \includegraphics[width=0.24\textwidth]{irfs/gs1_d.pdf}
%     \includegraphics[width=0.24\textwidth]{irfs/gs10_d.pdf}
%     \includegraphics[width=0.24\textwidth]{irfs/usbkeven5y_d.pdf}
%     \includegraphics[width=0.24\textwidth]{irfs/vixcls_d.pdf}
%     \includegraphics[width=0.24\textwidth]{irfs/DEXUSEU.pdf}
%     \includegraphics[width=0.24\textwidth]{irfs/bofaml_us_aa_oas_d.pdf}
%     \includegraphics[width=0.24\textwidth]{irfs/bofaml_us_bbb_oas_d.pdf}
%     \includegraphics[width=0.24\textwidth]{irfs/bofaml_us_hyld_oas_d.pdf}
%     \includegraphics[width=0.24\textwidth]{irfs/gold_eur_d.pdf}
%     \includegraphics[width=0.24\textwidth]{irfs/gold_usd_d.pdf}
%     %\includegraphics[width=0.24\textwidth]{irfs/oil_brent_d.pdf}
%     \caption{Impulse responses for US and international variables. The solid black line shows the posterior median estimate for the mixture model, alongside the 68 and 90 percent posterior credible sets in shades of blue. The red lines mark the posterior median, $16$/$84$th and $5$/$95$th percentile for the non-mixture specification.}
%     \label{fig:irf-USint}
% \end{figure}

\section{Conclusions}
Legal challenges to the conduct of the ECB's monetary policy have the potential to seriously impact policy making by reducing the variety of tools at the disposal of the central bank. This implies a substantial reduction in the independence and flexibility of the central bank. In this paper, we analyze three lawsuits surrounding the ECB's unconventional monetary policy programs. The number of corresponding ruling days is small and information on the precise timing of the announcements of the court decisions is not always available. Standard tools used in the literature have a difficult time in such a situation since the non-event days often include significant events which potentially jeopardize identification of the shock of interest. To cope with this, we develop an econometric model which discriminates between ruling and non-event days. With non-event days we refer to days that display very low levels of time series volatility and can thus safely be regarded as days without significant economic or political events.

We use this framework to analyze the effects of an expansionary ruling shock (which can be interpreted as an increase in central bank independence). The corresponding impulse responses are consistent with our prior expectations. Financial stress measures decrease and bond spreads relative to German Bunds decline; short-term interest rates, stock markets and inflation swaps increase. These effects are mostly driven by adjusting the expectations of financial market participants to the case where the ECB is left unrestricted in terms of its policy options, and maintains credibility in its conduct of monetary policy.

\small{\setstretch{1.1}
\addcontentsline{toc}{section}{References}
\bibliographystyle{custom.bst}
\bibliography{lit}}\normalsize\clearpage

\begin{appendices}
\begin{center}
    \begin{LARGE}
    \textbf{Appendix}
    \end{LARGE}
\end{center}

\setcounter{equation}{0}
\renewcommand\theequation{A.\arabic{equation}}
\section{Bayesian inference and sampling algorithm}\label{app:A:econometrics}
\subsection{Prior setup}
We assume independent inverse Gamma priors on the component-specific variances $\vartheta_j^2$ in a symmetric fashion for $j=1,\hdots,J$:
\begin{equation*}
    \vartheta_j^2\sim\mathcal{G}^{-1}(a_0,b_0),
\end{equation*}
with $a_0=b_0=0.1$ establishing a weakly informative setup. Moreover, we assume independent Gaussian priors centered on zero for the factor loadings:
\begin{equation*}
    \bm{\lambda} \sim\mathcal{N}(\bm{0}_M,\tau\bm{I}_M),
\end{equation*}
where $\tau$ governs the tightness of the prior. We again opt for a comparatively uninformative setup and choose $\tau=1$.

To impose shrinkage on the number of non-empty cluster components, we collect the component weights in a $J$-dimensional vector $\bm{w} = (w_1,\hdots,w_J)'$ and assume a symmetric Dirichlet prior:
\begin{equation*}
    \bm{w} \sim \text{Dir}(c_0,\hdots,c_0).
\end{equation*}
The hyperparameter $c_0$ governs how superfluous components of the mixture distributions are treated, and is thus closely related to the degree of shrinkage on the number of non-empty components. Following \citet{malsiner2016model}, we impose a Gamma prior on this parameter:
\begin{equation*}
    c_0 \sim \mathcal{G}(d,dJ).
\end{equation*}
Note that the expected value of $c_0$ is $1/J$ with variance $1/(dJ)^2$, which implies that both the level of shrinkage and tightness of the prior increases with the maximum number of components $J$. We set $d=10$ in our empirical application, emptying superfluous components aggressively.

\subsection{Posterior distributions}
\subsubsection{Sampling the component-specific variances}
Let $T_j$ denote the number of observations assigned to component $j$. The component-specific variances, conditional on $\delta_t$ for $t=1,\hdots,T$, take a textbook form and are inverse Gamma distributed:
\begin{align}
    \vartheta_j^2&\sim\mathcal{G}^{-1}(a_j,b_j),\label{eq:post-theta2}\\
    a_j &= a_0 + T_j/2,\nonumber\\
    b_j &= b_0 + \frac{1}{2}\sum_{t:\delta_t=j} (\bm{y}_{t} - d_t\bm{\lambda}f_t)'(\bm{y}_{t} - d_t\bm{\lambda}f_t).\nonumber
\end{align}

\subsubsection{Sampling parameters associated with the mixture model}
The conditional posterior of the component weights $\bm{w}$ is a Dirichlet distribution:
\begin{equation}
    \bm{w}\sim\text{Dir}(c_1,\hdots,c_J), \quad \text{with} \quad c_j = c_0 + T_j.\label{eq:post-w}
\end{equation}
The latent component-allocation indicators $\delta_t$ are sampled $t$-by-$t$ based on
\begin{equation}
    \tilde{\xi}_{jt} = \Pr(\delta_t=j|\bm{y}_t;\vartheta_j^2\bm{\Omega},\bm{w}) \propto w_j\mathcal{N}(\bm{y}_t|\bm{0}_M,\vartheta_j^2\bm{\Omega}).\label{eq:post-xi}
\end{equation}
The elements of the normalized probability vector $\bm{\xi}_t = (\xi_{1t},\hdots,\xi_{Jt})'$ are given by $\xi_{jt} = \tilde{\xi}_{jt} / \sum_{j=1}^J \tilde{\xi}_{jt}$. Recall that the non-event set is constructed by considering the component variances in ascending order, such that the first cluster has the lowest variance, $\vartheta_1^2$. Observations in the set $\mathcal{A}_{\text{R}}$ are always assigned to the first cluster.

Combining the likelihood with the prior for the intensity parameter $c_0$ using Bayes' theorem yields $p(c_0|\bm{w}) \propto p(\bm{w}|c_0) p(c_0)$ from which we sample using a Metropolis-Hastings algorithm with a log-Normal random walk proposal. 

Denote the current draw by $c_0^{(s)}$ and let $c_0^{(\ast)}$ be a proposal from the candidate density, $\log\left(c_0^{(\ast)}\right)\sim\mathcal{N}\left(\log\left(c_0^{(s)}\right),\varsigma^2\right)$. The parameter $\varsigma^2$ is a tuning parameter which we scale during half of the burnin-phase of the algorithm to yield an acceptance probability between 20 to 40 percent. The logarithm of the acceptance probability based on the likelihood ratio and the asymmetric candidate density is given by
\begin{align}
    \log(\zeta) &= \left(c_0^{(\ast)} - c_0^{(s)}\right) \left(\sum_{j=1}^J\sum_{t\in\mathcal{A}_{\text{NR}}} \log\left(\xi_{jt}\right)\right) + \log\left(\Gamma\left(Jc_0^{(\ast)}\right)\right) - \log\left(\Gamma\left(Jc_0^{(s)}\right)\right)\label{eq:MHacc}\\
    &- J\left[\log\left(\Gamma\left(c_0^{(\ast)}\right)\right) - \log\left(\Gamma\left(c_0^{(s)}\right)\right)\right] + (d-1) \left(\log\left(c_0^{(\ast)}\right) - \log\left(c_0^{(s)}\right)\right) \nonumber\\
    &- dJ \left(c_0^{(\ast)} - c_0^{(s)}\right) + \log\left(c_0^{(\ast)}\right) - \log\left(c_0^{(s)}\right),\nonumber
\end{align}
where $\Gamma(\bullet)$ denotes the Gamma function. The draw is accepted with probability $\min(\zeta,1)$, otherwise, the previous draw is retained.

\subsubsection{Sampling parameters associated with the factor model}
The factor loadings are estimated exclusively from observations in the set $\mathcal{A}_{\text{R}}$, and follow independent Gaussian posterior distributions for $i=1,\hdots,M$ with moments
\begin{align}
    \lambda_i &\sim \mathcal{N}\left(\overline{\lambda}_i,\overline{\Lambda}_i\right),\label{eq:post-lambda}\\
    \overline{\Lambda}_i &= \left(\sum_{t\in\mathcal{A}_{\text{R}}} \left(\frac{f_t}{\vartheta_1\omega_i}\right)^2 + \tau^{-1}\right)^{-1},\nonumber\\
    \overline{\lambda_i} &= \overline{\Lambda}_i\left(\sum_{t\in\mathcal{A}_{\text{R}}}\frac{y_{it} f_t}{\vartheta_1^2\omega_i^2}\right).\nonumber
\end{align}
The factors are sampled for $t\in\mathcal{A}_\text{R}$ from a Gaussian distribution with moments given by 
\begin{align}
    f_t & \sim\mathcal{N}\left(\overline{f}_t,\overline{F}_t\right),\label{eq:post-ft}\\
    \overline{F}_t &= \left(\vartheta_1^{-2}\bm{\lambda}'\bm{\Omega}^{-1}\bm{\lambda} + 1\right)^{-1},\nonumber\\
    \overline{f}_t &= \overline{F}_t \left(\vartheta_1^{-2}\bm{\lambda}'\bm{\Omega}^{-1}\bm{y}_t\right).\nonumber
\end{align}

\subsection{Sampling algorithm}
The posterior distributions listed in the previous subsection can be used to set up a Gibbs sampling algorithm that samples iteratively from these conditionals. After initializing all variables, our sampler cycles through the following steps:
\begin{enumerate}
    \item Sample the cluster-specific variances from their inverse Gamma conditional posterior distribution given by Eq. (\ref{eq:post-theta2}).
    \item Sample the latent cluster-indicators $\delta_t$ using the respective probabilities given by Eq. (\ref{eq:post-xi}).
    \item Sample the component weights $\bm{w}$ from a Dirichlet distribution with moments stated in Eq. (\ref{eq:post-w}). After this step, the variances and all other parameters associated with the mixture model are permuted such that they are in ascending order and $j=1$ is the cluster with minimal variance $\vartheta_1^2$.
    \item Sample the intensity parameter $c_0$ using the Metropolis-Hastings step with the acceptance probability given by Eq. (\ref{eq:MHacc}).
    \item Sample the factor loadings $\bm{\lambda}$ from their Gaussian conditional posterior distribution with moments given in Eq. (\ref{eq:post-lambda}) for $i=1,\hdots,M$.
    \item Sample the factor $f_t$ on a $t$-by-$t$ basis for the ruling set $\mathcal{A}_{\text{R}}$ from their Gaussian conditional posterior distribution stated in Eq. (\ref{eq:post-ft}). This step involves checking the sign restriction of the factor. Pre-multiply $\bm{\lambda}$ and $f_t$ by negative one if the restriction is not fulfilled, otherwise omit this step.
\end{enumerate}
In the empirical application of the paper, we produce a set of $3,000$ draws from the desired full conditional posterior after disregarding the initial $2,000$ draws as burnin. To reduce autocorrelation in our MCMC sample, we consider each third from the retained draws and perform inference based on the remaining $1,000$ draws. 

\clearpage
\section{Data}\label{app:B:data}
\setcounter{table}{0}
\renewcommand\thetable{B.\arabic{table}}

{Table \ref{tab:variables} provides a list of variables alongside their source and scale. We obtain data from the statistical data warehouse of the European Central Bank (ECB SDW, \href{https://sdw.ecb.europa.eu}{sdw.ecb.europa.eu}) and the database of the Federal Reserve Bank of St. Louis (FRED, \href{https://fred.stlouisfed.org}{fred.stlouisfed.org}), Yahoo Finance and Macrobond. If applicable, maturities are indicated with ``m'' for months and ``yr'' for years. Scale refers to the underlying transformation of the series and how impulse response functions are to be interpreted: basis points (BPs), in percent (\%) or no transformation (raw). EUR refers to euros and USD denote US dollars. Country codes are: Austria (AT), Belgium (BE), Germany (DE), Spain (ES), Finland (FI), France (FR), Greece (GR), Ireland (IE), Italy (IT), the Netherlands (NL) and Portugal (PT).}

\begin{table}[ht]
\caption{List of variables.}\label{tab:variables}
\centering\small
\begin{tabular}{llcc}
  \toprule
\textbf{Label} & \textbf{Information} & \textbf{Scale} & \textbf{Source} \\ 
  \midrule
  OIS $D$ & Overnight index swap interest rates & BPs & ECB SDW \\ 
         & $D\in\{$1-m, 3-m, 6-m, 1-yr, 2-yr$\}$ &  &  \\ 
  Infl. swap $D$ & Inflation swaps (EUR) & BPs &  \\ 
                & $D\in\{$1-yr, 2-yr, 3-yr, 4-yr, 5-yr, 10-yr$\}$ & &  \\ 
  CISS ($C$) & Composite indicator of systemic stress & raw &  \\ 
             & $C\in\{$AT, BE, DE, ES, FI, FR, IE, IT, NL, PT$\}$ &  &  \\ 
  EuroStoxx 50 & ESTOXX50 index & \% &  \\ 
  \midrule
  Hyld OAS (EA) & ICE BofA EA High Yield Option-Adjusted Spread & BPs & FRED \\ 
  AA OAS (EA) & ICE BofA EA AA Option-Adjusted Spread & BPs &  \\ 
  BBB OAS (EA) & ICE BofA EA BBB Option-Adjusted Spread & BPs &  \\ 
  Hyld OAS (US) & ICE BofA US High Yield Option-Adjusted Spread & BPs &  \\ 
  AA OAS (US) & ICE BofA US AA Option-Adjusted Spread & BPs &  \\ 
  BBB OAS (US) & ICE BofA US BBB Option-Adjusted Spread & BPs &  \\ 
  GS 1y & US 1-yr treasury yield & BPs &  \\ 
  GS 10y & US 10-yr treasury yield & BPs &  \\ 
  BE Infl. 5y (US) & US break even inflation & BPs &  \\ 
  FX USD/EUR & USD/EUR exchange rate & \% &  \\ 
  NASDAQ & NASDAQ composite index & \% &  \\ 
  VIX & CBOE Volatility Index & \% &  \\ 
  Oil prices & Crude oil prices: Brent (Europe) & \% &  \\ 
  Gold (USD) & Gold fixing price London Bullion Market (in USD) & \% &  \\ 
  Gold (EUR) & Gold fixing price London Bullion Market (in EUR) & \% &  \\ 
  \midrule
  Stock market ($C$) & Stock index for $C$ & \% & Yahoo Finance \\ 
                     & $C\in\{$AT, BE, DE, ES, FI, FR, GR, IE, IT, NL, PT$\}$ &  &  \\ 
                     & \textit{ATX} (AT), \textit{BEL 20} (BE), \textit{DAX} (DE), \textit{IBEX 35} (ES) &  &  \\ 
                     & \textit{OMXH 25} (FI), \textit{CAC 40} (FR), \textit{Athex 20} (GR) &  &  \\ 
                     & \textit{ISEQ 20} (IE), \textit{FTSE MIB} (IT), \textit{AEX} (NL)&  &  \\ 
                     & \textit{PSI 20} (PT) &  &  \\ 
  \midrule
  Spread ($C$/DE) & 10-yr government bond yield of $C$ minus DE & BPs & Macrobond \\ 
                  & $C\in\{$AT, BE, ES, FI, FR, GR, IE, IT, NL, PT$\}$ &  &  \\ 
   \bottomrule
\end{tabular}
\end{table}

\clearpage
\section{Additional empirical results}\label{app:C:results}
\setcounter{figure}{0}
\renewcommand\thefigure{C.\arabic{figure}}

\begin{figure}[!h]
    \includegraphics[width=\textwidth]{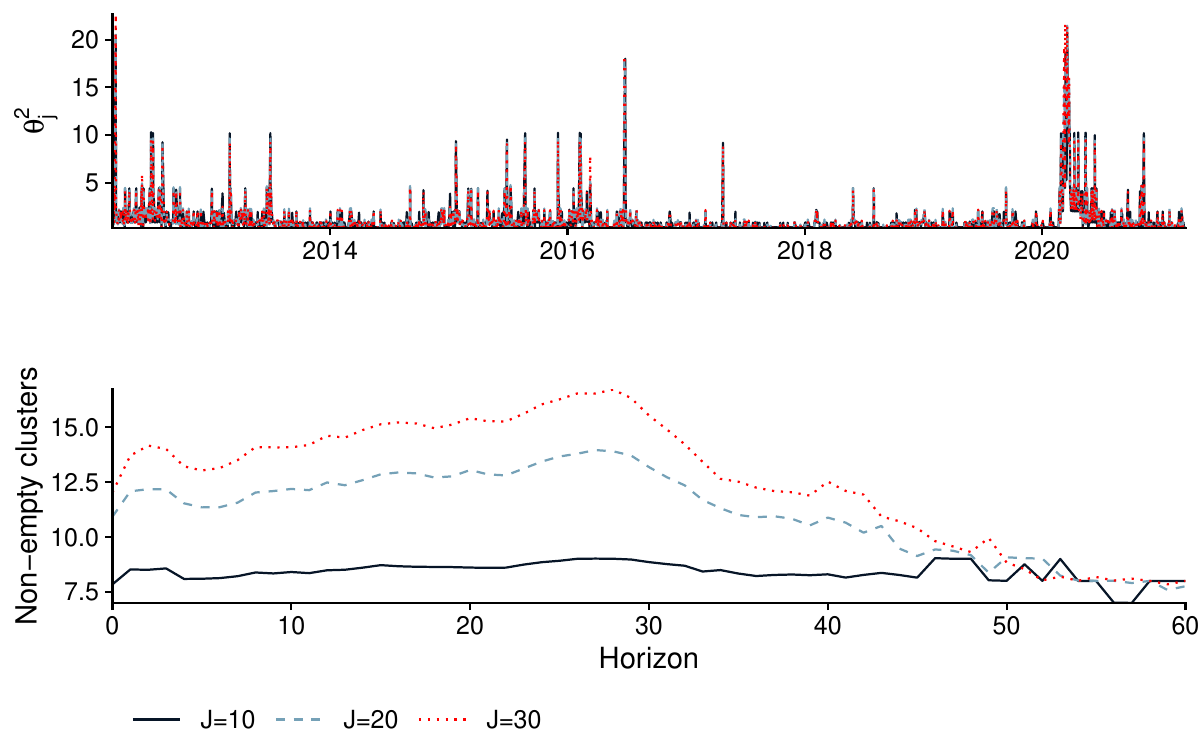}
    \caption{Variance clusters (posterior median for $h=0$) and number of non-empty components (posterior mean) across different specifications of the maximum number of components $J$ to determine the non-event set.}
    \label{fig:app-components}
\end{figure}

{Figure \ref{fig:app-components} reports estimates of the common volatilities for varying numbers of maximum mixture components $J$ and $h=0$ in the upper panel. The lower panel indicates the posterior mean estimate of the number of non-empty clusters. While assuming a larger number of clusters a priori results in a larger estimate of non-empty clusters for shorter horizons $h$, the estimated variance paths look very similar. The specific number of a priori clusters $J$ has virtually no impact on the estimated impulse response functions.}

\end{appendices}
\end{document}